\documentclass[a4paper,11pt]{article}

\usepackage{jheppub}
\usepackage[table]{xcolor}
\usepackage{slashbox}
\allowdisplaybreaks
\usepackage{graphicx}
\usepackage{epstopdf}

\newcommand{\be}{\begin{equation}}
\newcommand{\ee}{\end{equation}}
\newcommand{\bea}{\begin{eqnarray}}
\newcommand{\eea}{\end{eqnarray}}
\newcommand{\nn}{\nonumber}

\newcommand{\bdm}{\begin{displaymath}}
\newcommand{\edm}{\end{displaymath}}

\title{Gravitational cubic-in-spin interaction at the
\\next-to-leading post-Newtonian order}

\author[a,b]{Mich\`ele Levi,}
\author[b]{Stavros Mougiakakos,}
\author[a]{and Mariana Vieira}

\affiliation[a]{Niels Bohr International Academy, Niels Bohr Institute,
University of Copenhagen,
\\Blegdamsvej 17, 2100 Copenhagen, Denmark}
\affiliation[b]{Institut de Physique Th\'eorique,
CEA \& CNRS, Universit\'e Paris-Saclay,
\\ 91191 Gif-sur-Yvette, France}

\emailAdd{michelelevi@nbi.ku.dk, stavros.mougiakakos@ipht.fr,dcp182@alumni.ku.dk}

\abstract{
In this work we derive for the first time the complete gravitational 
cubic-in-spin effective action at the next-to-leading order in the 
post-Newtonian (PN) expansion for the interaction of generic compact 
binaries via the effective field theory for gravitating spinning objects, 
which we extend in this work. This sector, which enters at the fourth 
and a half PN (4.5PN) order for rapidly-rotating compact objects, 
completes finite-size effects up to this PN order, and is the first sector 
completed beyond the current state of the art for generic compact binary 
dynamics at the 4PN order. At this order in spins with gravitational 
nonlinearities we have to take into account additional terms, which arise 
from a new type of worldline couplings, due to the fact that at this order 
the Tulczyjew gauge for the rotational degrees of freedom, which involves 
the linear momentum, can no longer be approximated only in terms of the 
four-velocity. One of the main motivations for us to tackle this sector 
is also to see what happens when we go to a sector, which corresponds to 
the gravitational Compton scattering with quantum spins larger than one, 
and maybe possibly also get an insight on the inability to uniquely fix 
its amplitude from factorization when spins larger than two are 
involved. A general observation that we can clearly make already is that 
even-parity sectors in the order of the spin are easier to handle than odd 
ones. In the quantum context this corresponds to the greater ease of 
dealing with bosons compared to fermions.}

\begin{document}


\maketitle

\flushbottom

\section{Introduction} 
\label{intro}

Since the first detection of gravitational waves (GWs) from a binary black 
hole coalescence was announced in 2016 \cite{Abbott:2016blz}, 
it has become increasingly pressing 
to provide high-precision theoretical predictions for the modeling of GW 
templates. The latter rely on implementing analytical 
results obtained within the post-Newtonian (PN) approximation of classical 
Gravity \cite{Blanchet:2013haa} via the Effective-One-Body approach 
\cite{Buonanno:1998gg}. In particular in recent years we have 
made a remarkable progress in pushing the precision frontier for the 
orbital dynamics of compact binaries, i.e.~whose components are generic 
compact objects, such as black holes or neutron stars. The complete 
state of the art to date for the orbital dynamics of 
a generic compact binary is shown in table \ref{stateoftheart}. 

As a measure for the loop computational scale we show in table 
\ref{stateoftheart} the number of $n$-loop graphs that enter at the
N$^n$LO in $l$ powers of the spin, i.e.~up to the $l$th spin-induced 
multipole moment, in the sectors approached to date. The count is based on 
computations carried out with the effective field theory (EFT) of PN 
Gravity \cite{Goldberger:2004jt}, which use the Kaluza-Klein decomposition 
of the field from \cite{Kol:2007bc}, that has considerably facilitated 
high-precision computations within the EFT approach
\cite{Kol:2007bc,Levi:2008nh,Gilmore:2008gq,Levi:2010zu,Foffa:2011ub,
Levi:2011eq,Levi:2014gsa,Levi:2015msa,Levi:2015uxa,Levi:2015ixa,
Foffa:2016rgu,Foffa:2019hrb,Blumlein:2019zku,Levi:2020kvb,Levi:2020uwu,
Levi:2020lfn}. As can be seen the current complete state of the art is at 
the 4PN order, whereas the next-to-leading order (NLO) cubic-in-spin 
sector that enters at the 4.5PN order for maximally-rotating objects is 
evaluated in this paper. 
All of the sectors at the current state of the art (but the top right
entry at the 4PN order for the non-rotating case) are available in the 
public \texttt{EFTofPNG} code at 
\url{https://github.com/miche-levi/pncbc-eftofpng} \cite{Levi:2017kzq}. 

\begin{table}[t]
\begin{center}
\begin{tabular}{|l|r|r|r|r|r}
\hline
\backslashbox{\quad\boldmath{$l$}}{\boldmath{$n$}} 
&  (N\boldmath{$^{0}$})LO
& N\boldmath{$^{(1)}$}LO & \boldmath{N$^2$LO}
& \boldmath{N$^3$LO} & \boldmath{N$^4$LO}\\
\hline
\boldmath{S$^0$} & 1 & 0  & 3 
& 0  & 25 \\
\hline
\boldmath{S$^1$} & 2 & 7  & 32 & 174
& \\
\hline
\boldmath{S$^2$} & 2 & 2  & 18 
& 52 & \\ 
\hline
\boldmath{S$^3$} & 4  & \cellcolor[gray]{0.9} 24
& \cellcolor[gray]{0.9} & \cellcolor[gray]{0.9} & \cellcolor[gray]{0.9}\\
\hline
\boldmath{S$^4$} & 3& \cellcolor[gray]{0.9} 5 & \cellcolor[gray]{0.9} 
& \cellcolor[gray]{0.9} & \cellcolor[gray]{0.9}\\
\end{tabular}
\caption{The complete state of the art of PN gravity theory for the 
orbital dynamics of generic compact binaries. Each PN correction enters at 
the order $n + l + \text{Parity}(l)/2$, where the parity is $0$ or $1$ for 
even or odd $l$, respectively. We elaborate on the meaning of the 
numerical entries and the gray area in the text.} 
\label{stateoftheart}
\end{center}
\end{table}

Let us stress that in order to attain a certain level of PN accuracy, the 
various sectors should be tackled across the diagonals of table 
\ref{stateoftheart}, rather than along the axes, namely progress must be 
made by going in parallel to higher loops and to higher orders of the 
spin. In general, the former involves more computational challenges of 
loop technology and tackling associated divergences, whereas the latter 
necessitates an improvement of the fundamental understanding of spin in 
gravity, and tackling finite-size effects with spin \cite{Levi:2018nxp}. 
The latter enter first at the 2PN order \cite{Barker:1975ae} from the LO 
spin-induced quadrupole. Within the EFT approach, whose extension to the 
spinning case was first approached in \cite{Porto:2005ac}, finite-size 
effects include as additional parameters the Wilson coefficients, that 
correspond, e.g., to the multipole deformations of the object due to 
its spin, as in \cite{Poisson:1997ha} for the spin-induced quadrupole.

With a considerable time gap from the LO result, the NLO spin-squared 
interaction was treated in a series of works 
\cite{Porto:2008jj,Steinhoff:2008ji,Hergt:2008jn,Hergt:2010pa,
Levi:2015msa}, where in \cite{Levi:2015msa} it was derived within 
the formulation of the EFT for gravitating spinning objects, that provided 
the leading non-minimal couplings to all orders in spin. 
The LO cubic- and quartic-in-spin interactions were first tackled 
in \cite{Hergt:2007ha,Hergt:2008jn} for black holes. In 
\cite{Levi:2014gsa}, based on the formulation presented in 
\cite{Levi:2015msa}, these were derived for generic compact objects, where 
also the quartic-in-spin interaction was completed. Only specific pieces 
of the latter were recovered in \cite{Vaidya:2014kza} via S-matrix 
combined with EFT techniques, whereas \cite{Marsat:2014xea}, which treated 
only cubic-in-spin effects, also provided the LO effects in the energy 
flux. Following the work in \cite{Levi:2014gsa,Levi:2015msa}, the case of 
black holes was then also completed for the LO sectors to all orders in spin
\cite{Vines:2016qwa}. Finally, the NNLO spin-squared interaction was derived in 
\cite{Levi:2015ixa}. Notably the latter results together with the complete 
quartic-in-spin results for generic compact objects in \cite{Levi:2014gsa}, 
both at the 4PN order, were derived exclusively within the EFT formulation 
of spinning gravitating objects \cite{Levi:2015msa}. Building on the latter, 
recent works further pushed at the 4.5PN \cite{Levi:2020kvb} and 5PN 
\cite{Levi:2020uwu,Levi:2020lfn} orders for maximally-spinning objects.

\begin{figure}[t]
\centering
\includegraphics[scale=1]{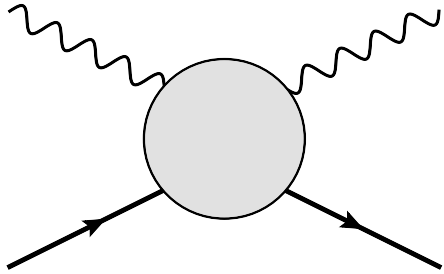}
\caption{The gravitational Compton scattering relevant as of the one-loop 
level. The gravitational Compton amplitude involves two massive spinning 
particles and two gravitons, where factorization constraints do 
not uniquely determine the amplitude for $s>2$ 
\cite{Arkani-Hamed:2017jhn}.}
\label{compton}
\end{figure}

Recently, there has also been a surge of interest in harnessing modern 
advances in scattering amplitudes to the problem of a coalescence of a 
compact binary. Notably, a new implementation for the non-rotating case 
to the derivation of classical potentials was carried out in 
\cite{Bern:2019nnu,Bern:2019crd}. Further, based on a new spinor-helicity
formalism introduced in \cite{Arkani-Hamed:2017jhn} for massive particles 
of any spin, new approaches to the computation of spin effects of black 
holes in the classical potential were put forward in 
\cite{Cachazo:2017jef,Guevara:2017csg} and then in 
\cite{Chung:2018kqs,Chung:2019duq}. In these approaches classical effects 
with spin to the $l$th order correspond to scattering amplitudes involving 
a quantum spin of $s=l/2$. In particular as of the one-loop level the 
gravitational Compton amplitude in figure \ref{compton} is required, where 
factorization constraints can not uniquely determine the amplitude for 
$s>2$ \cite{Arkani-Hamed:2017jhn}. The gray area in table 
\ref{stateoftheart} then corresponds in the quantum context to where the 
gravitational Compton scattering with a spin $s > 1$ is required.

Notably, the gray area in table \ref{stateoftheart} also corresponds to, 
as was already pointed out in \cite{Levi:2015msa}, where we can no longer 
take the linear momentum $p_\mu$, with which the generic formulation in 
\cite{Levi:2015msa} was made, to be its leading approximation given by 
$m \tfrac{u^{\mu}}{\sqrt{u^2}}$, as in all previously tackled spin sectors, 
but we have to take into account corrections to the linear 
momentum from the non-minimal coupling part of the spinning particle 
action. Can we then get a well-defined result? Can we get an insight from 
examining this new feature at the classical level on the non-uniqueness of 
the graviton Compton amplitude with $s > 2$?

This work builds on the formalism of the EFT for gravitating spinning  
objects introduced in \cite{Levi:2015msa} and the implementation on 
\cite{Levi:2014gsa}, to compute the cubic-in-spin interaction at the 
NLO, that enters at the 4.5PN order for maximally-rotating compact 
objects, pushing the current state of the art of PN theory in general and 
with spins in particular \cite{Levi:2016ofk}, and that is the leading 
sector in the intriguing gray area of table \ref{stateoftheart}. 
We compute the complete sector, taking into account all interactions that 
include all possible spin multipoles terms up to and including the octupole. 
Thus beyond pushing the state of the art in PN theory, there are two 
conceptual objectives that we get to address in this work: 1.~To learn how the 
spin dependence of the linear momentum affects the results in the interaction; 
2.~To see whether this new feature is related with the non-uniqueness of 
the gravitational Compton amplitude of higher-spin states, or get any 
possible insight on this non-uniqueness.

The paper is organized as follows. In section \ref{formulation} we go 
over the formulation from \cite{Levi:2015msa}, and the necessary 
ingredients to evaluate the more familiar part of the sector. In section 
\ref{Feyncompute} we present the essential computation, where the linear 
momentum assumes its leading approximation in terms of the four-velocity, 
as in all past evaluations of spin sectors. In section \ref{newfromgauge?} 
we find the new contributions arising from the spin-dependent correction 
to the linear momentum, which matters as of this sector, and gives rise to 
a new type of worldline-graviton coupling. In section \ref{finalaction} we 
assemble the final action of this sector, and finally we conclude in section 
\ref{lafin} with some observations and questions.


\section{The EFT of gravitating spinning objects}
\label{formulation}

Let us consider the ingredients of the theory that are required in order to 
carry out the evaluation of this sector, that contains spins up to cubic order 
along with first gravitational nonlinearities. This evaluation will build on 
the EFT of gravitating spinning objects formulated in \cite{Levi:2015msa}, 
and its implementation from LO up to the state of the art at the 4PN order 
in \cite{Levi:2015msa,Levi:2015uxa,Levi:2015ixa,Levi:2014gsa,Levi:2016ofk}. 
We will also use here the Kaluza-Klein decomposition of the field 
\cite{Kol:2007bc,Kol:2010ze}, which was adopted in all high-order PN 
computations both with and without spins for its facilitating virtues 
\cite{Levi:2018nxp}, and follow conventions consistent with the 
abovementioned works. 
Further, we keep similar gauge choices, notational and pictorial 
conventions as presented in \cite{Levi:2015msa}.

The effective action we start from is that of a two-particle system
\cite{Levi:2018nxp}, with each of the particles described by the 
effective point-particle action of a spinning particle, that was provided in 
\cite{Levi:2015msa}. This effective action contains a purely gravitational 
piece, from which the propagators and self-interacting vertices are 
derived. The Feynman rules for the propagator and the time insertions on 
the propagators are given, e.g.~in eqs.~(5)-(10) of \cite{Levi:2011eq}, and 
for the cubic gravitational vertices in eqs.~(2.10)-(2.13), and (2.15) of 
\cite{Levi:2015uxa}. Further, for each of the two particles, the worldline 
action of a spinning particle is considered from \cite{Levi:2015msa}, 
where its non-minimal coupling spin-induced part was constructed to all 
orders in spin, and then gauge freedom of the rotational DOFs is incorporated 
into the action. 

We recall that this action has the form
\begin{align} \label{spinact}
S_{\text{pp}}(\sigma)=&\int d\sigma\left[-m \sqrt{u^2}
-\frac{1}{2} \hat{S}_{\mu\nu} \hat{\Omega}^{\mu\nu}
-\frac{\hat{S}^{\mu\nu} p_{\nu}}{p^2} 
\frac{D p_{\mu}}{D \sigma} + L_{\text{SI}}\right],
\end{align}
given in terms of the four-velocity $u^{\mu}$, the linear momentum $p_\mu$, 
and the rotational DOFs in a generic gauge, denoted with a hat, 
e.g.~$\hat{S}_{\mu\nu}$,
and the label ``SI'' stands for the spin-induced non-minimal coupling 
part of the action. For the sector evaluated here the latter part will 
consist of its two leading terms given by
\begin{align} \label{nmc}
L_{\text{SI}} =&
-\frac{C_{ES^2}}{2m} \frac{E_{\mu\nu}}{\sqrt{u^2}} S^\mu S^\nu
-\frac{C_{BS^3}}{6m^2}D_{\lambda}\frac{B_{\mu\nu}}{\sqrt{u^2}}
S^{\mu} S^{\nu}S^\lambda,
\end{align}
in which we use the definite-parity curvature components defined as
\bea
E_{\mu\nu}&\equiv& R_{\mu\alpha\nu\beta}u^{\alpha}u^{\beta}, \label{eq:E}\\
B_{\mu\nu}&\equiv& \frac{1}{2} \epsilon_{\alpha\beta\gamma\mu} 
R^{\alpha\beta}_{\,\,\,\,\,\,\,\delta\nu}u^{\gamma}u^{\delta}\label{eq:B},
\eea
for the electric and magnetic components of even and odd parity, respectively.
Notice also that here we need to use the Levi-Civita tensor density in curved spacetime, 
$\epsilon_{\alpha\beta\gamma\lambda}=\sqrt{-g}\,e_{\alpha\beta\gamma\lambda}$, 
in which $g$ is the determinant of $g_{\mu\nu}$, and $e_{\alpha\beta\gamma\lambda}$ 
is the totally antisymmetric Levi-Civita symbol with $e_{0123}=+1$.
We note also that we use here a classical version of the Pauli-Lubanski pseudovector, 
$S^{\mu}$, as first defined in \cite{Levi:2014gsa}
\begin{align}\label{svec}
S_{\mu}=\frac{1}{2}\epsilon_{\alpha\beta\gamma\mu}S^{\alpha\beta}\frac{p^{\gamma}}{p},
\end{align}
which is with a reverse sign with respect to the definition in \cite{Levi:2015msa}, 
that was applied up to the quadratic-in-spin order, where this sign choice does 
not matter. The spin tensor that is used in eq.~\eqref{svec} is related to the spin 
in the generic gauge $\hat{S}_{\mu\nu}$ via the projection of the latter onto the 
spatial hypersurface of the rest frame according to eq.~(3.29) in 
\cite{Levi:2015msa}. 

We recall that in 
eq.~\eqref{spinact} there is an extra term, which appears in the action 
from the restoration of gauge freedom of the rotational DOFs. This term, 
which is essentially the Thomas precession, as discussed in detail in
\cite{Levi:2015msa} (and recovered recently as ``Hilbert space matching'' 
in \cite{Guevara:2018wpp,Chung:2019duq}), contributes to all orders in the 
spin as of the LO spin-orbit sector, and in particular also to finite-size 
spin effects, though it does not encode any UV physics, but rather in the 
context of an effective action, just accounts for the fact that a 
relativistic gravitating object has an extended measure.

Since we compute here the \textit{complete} NLO cubic-in-spin sector our 
graphs will contain all multipoles in the presence of spin up to the 
spin-induced octupole, i.e.~also including the mass, spin and spin-induced 
quadrupole. For this reason we need to use Feynman rules of 
worldline-graviton coupling to NLO for all of these multipoles, where in 
this work we need to derive further new rules for the octupole couplings. 
The Feynman rules required for the mass couplings are given in eqs.~(64), 
(67), (68), (79), (81), (93), (95) of \cite{Levi:2010zu}.
Next, we approach the Feynman rules linear in spin, noting that first we 
have kinematic contributions as noted in eq.~(5.28) of 
\cite{Levi:2015msa}, that are linear in the spin but have no field 
coupling, which we will take into account in section \ref{finalaction}.
The Feynman rules required for the linear-in-spin couplings are given in 
eqs.~(5.29)-(5.31) of \cite{Levi:2015msa}, and eqs.~(2.31)-(2.34) of 
\cite{Levi:2015uxa}. For the spin quadrupole couplings the rules are given 
in eqs.~(2.18)-(2.24) of \cite{Levi:2015ixa}, and for the LO spin octupole 
couplings they are found in eqs.~(2.19),(2.20) of \cite{Levi:2014gsa}.

As we noted in addition to the abovementioned Feynman rules, further rules 
are required here for the spin-induced octupole worldline-graviton 
coupling. The two Feynman rules of the scalar and vector components of the 
KK decomposition, which appeared already at LO in \cite{Levi:2014gsa} 
should be extended to a higher PN order, and further we will have new rules
that enter for the one-graviton coupling of the tensor component of the KK 
fields, and a couple of two-graviton couplings, involving again the scalar 
and the vector components of the KK fields.

The extended rules for the one-graviton couplings are then given as
follows:
\begin{align}
\label{eq:s3A}  \parbox{12mm}{\includegraphics[scale=0.6]{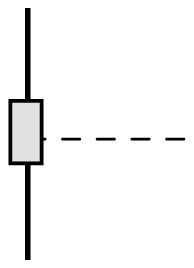}}
 =& \int dt \,\, 
 \Bigg[\frac{C_{\text{BS}^3}}{12m^2} S_{i}S_{j}\epsilon_{klm}
 \Big[A_{k,ijl} \Big(S_{m} \Big(1+\frac{1}{2}v^2\Big)-\frac{1}{2}v^mS_nv^n \Big)
 \nn\\
& \qquad \qquad 
+ S_m \Big( v^lv^n \big(A_{i,njk}-A_{n,ijk}\big)
+ v^l\big(\partial_tA_{k,ij}+\partial_tA_{i,jk}\big)
+ v^i\partial_tA_{k,jl} \Big) \Big] \Bigg]
,\\
\label{eq:s3phi}   \parbox{12mm}{\includegraphics[scale=0.6]{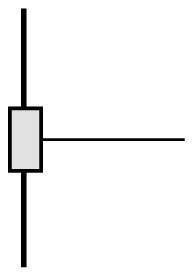}}
 =& \int dt \,\,\Bigg[\frac{C_{\text{BS}^3}}{3m^2}
 S_{i}S_{j} \epsilon_{klm}S_{m}v^l
 \Big( \phi_{,ijk}\,\Big(1+\frac{v^2}{2}\Big) 
 + v^i \partial_t \phi_{,jk} \Big) \Bigg],
\end{align}
in which the rectangular boxes represent the spin-induced octupole.

The new Feynman rules required here are given as follows:
\begin{align}
\label{eq:s3sigma} 
\parbox{12mm}{\includegraphics[scale=0.6]{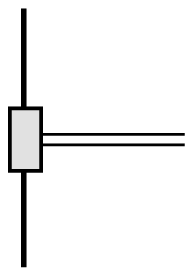}}
 =&\int dt \,\,\Bigg[\frac{C_{\text{BS}^3}}{12m^2}
  S_{i}S_{j}\epsilon_{klm}S_{m}\partial_i\partial_l
  \Big(\big(\partial_j\sigma_{k n} - \partial_n\sigma_{jk} \big) v^n
  -\partial_t \sigma_{jk}\Big) \Bigg],
\end{align}
for the one-graviton coupling, whereas for the two-graviton couplings we get:
\begin{align}
\label{eq:s3phiA}  \parbox{12mm}{\includegraphics[scale=0.6]{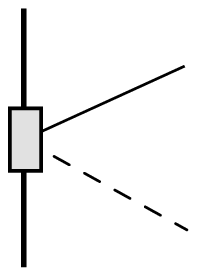}}
 =& \int dt \,\, \Bigg[ \frac{C_{\text{BS}^3}}{12m^2} S_{i}S_{j}\epsilon_{klm}S_{m}
\Big(6\phi A_{k,ijl} + 9\phi_{,i} A_{k,jl} 
+ 3\phi_{,k}\partial_j\big(A_{i,l} - A_{l,i} \big)
\nn\\
 &\qquad \qquad \qquad \qquad \qquad 
 + 4\phi_{,ij}A_{k,l} + 4\phi_{,jk} \big(A_{i,l} - A_{l,i}\big)
  +\delta_{ij}\phi_{,n}A_{l,kn}\Big)\Bigg],\\
   \label{eq:s3phiphi}   
   \parbox{12mm}{\includegraphics[scale=0.6]{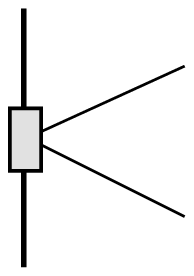}}
 =& \int dt \,\,\Bigg[\frac{C_{\text{BS}^3}}{3m^2} 
 S_{i}S_j\epsilon_{klm}S_{m} 
 \Big[ v^l 
 \Big( 2\phi_{,ijk} \phi + 3\phi_{,ij} \phi_{,k} 
 + 5\phi_{,i} \phi_{,jk} - \delta_{ij}\phi_{,n}\phi_{,kn} \Big)\nn\\
 &
\qquad \qquad \qquad \qquad \qquad \qquad 
\qquad \qquad \qquad \qquad \qquad \qquad \quad
+v^i \phi_{,lj}\phi_{,k}\Big] \Bigg].
\end{align}
We note that in these rules the spin is already fixed to the canonical gauge in 
the local frame, and all indices are Euclidean. Notice the complexity of these 
couplings with respect to other worldline couplings at the NLO level, 
and notice also the dominant role that the gravitomagnetic vector plays in 
the coupling to the odd-parity octupole, similar to the situation in the 
coupling to the spin dipole.
This is the first sector which necessitates to actually take the curved 
Levi-Civita tensor and the covariant derivative into account.

For this sector there is no need to extend the non-minimal coupling part 
of the spinning particle action and add higher dimensional operators 
beyond what was provided in \cite{Levi:2015msa}, but we need to pay 
special attention to the new feature that differentiates this specific 
sector from all the spin sectors which were tackled in the past. In this 
sector it is no longer sufficient to use the leading approximation for the 
linear momentum $p_\mu$ in terms of the four-velocity $u_\nu$ all 
throughout, rather one has to take into account the subleading
term in the linear momentum, which is linear in Riemann and quadratic in 
the spin, and becomes relevant first once we get to the level that is 
cubic in the spins and non-linear in gravity, i.e.~at this sector, 
as was already explicitly noted in \cite{Levi:2015msa}.
We will address in detail the particular contributions coming from this 
new feature in section \ref{newfromgauge?} below, after we have done in the 
following section the essential computation, which requires only the leading 
approximation to the linear momentum that is independent of spin, similar to 
what was considered in all past PN computations in spin sectors.

\section{The essential computation}
\label{Feyncompute}

In this section we carry out the perturbative expansion of the effective
action in terms of Feynman graphs, and provide the value of each diagram, 
while using the leading approximation of the linear momentum.
At the NLO level, i.e.~up to the $G^2$ order, all of the three
relevant topologies are realized with spins, even when the beneficial KK
decomposition of the field is used, as discussed in
\cite{Levi:2008nh,Levi:2010zu,Levi:2015msa,Levi:2018nxp}. As shown in
figures \ref{cubspin1g0loop}-\ref{cubspin1loop} below (drawn using 
Jaxodraw \cite{Binosi:2003yf,Binosi:2008ig} based on 
\cite{Vermaseren:1994je}) there is a total of $49=10+15+24$ graphs making
up this part of the sector, distributed among the relevant topologies of 
one- and two-graviton exchanges and cubic self-interaction, respectively. 
As shown in table \ref{stateoftheart} about half of the total graphs 
require a one-loop evaluation (the highest loop in this sector).
We note that as we go into the nonlinear part of the sector, the
options for the make-up of the interaction become more intricate.

At the one-graviton exchange level we only have two kinds of interaction
contributing, similar to the LO in \cite{Levi:2014gsa}, namely either an
octupole-monopole or a quadrupole-dipole interaction.
As noted in \cite{Levi:2014gsa} there are nice analogies among these
interactions according to the parity of the multipole moments involved.
Following these analogies the relevant graphs of one-graviton exchange
are easily constructed.
Yet, once we proceed to the level nonlinear in the gravitons further
types of interactions emerge. In particular, there are also interactions
involving various multipoles on two different points of the worldline, 
which add up to interactions that are cubic in the spin, such as a spin 
and a spin-induced quadrupole or two spin dipoles, on the same 
worldline, which can already be seen as of the NLO spin-squared sector
\cite{Levi:2015msa,Levi:2015ixa}.

We note that all the graphs in this sector should be included together
with their mirror images, i.e.~with the worldline labels
$1\leftrightarrow2$ exchanged.
For more specific details on the generation of the Feynman graphs, and
their evaluation, including the conventions and notations used here, we
refer the reader to \cite{Levi:2018nxp} and references therein. We note 
that the generation and the evaluation of the graphs was crosschecked 
using the publicly-available \texttt{EFTofPNG} code \cite{Levi:2017kzq}.


\subsection{One-graviton exchange}

\begin{figure}[t]
\centering
\includegraphics[width=\textwidth]{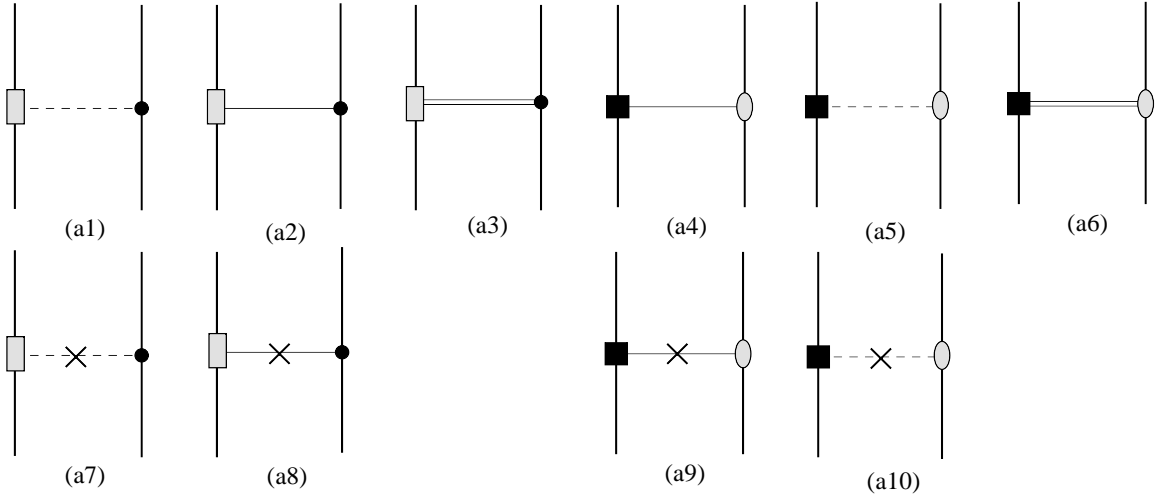}
\caption{The Feynman graphs of one-graviton exchange, which contribute to
the NLO cubic-in-spin interaction at the 4.5PN order for maximally
rotating compact objects. 
The graphs should be included together with their mirror images, 
i.e.~with the worldline labels $1\leftrightarrow2$ exchanged.
At the linear level of one-graviton exchange we only have two kinds of 
interactions contributing, similar to the LO in \cite{Levi:2014gsa}, namely 
either a quadrupole-dipole or an octupole-monopole interaction. As noted in
\cite{Levi:2014gsa} there are nice analogies among these interactions
according to the parity of the multipole moments involved. Following
these analogies the relevant graphs here are easily constructed.
Notice that we have here the four graphs that appeared at the LO with the
quadratic time insertions on the propagators at graphs (a7)-(a10), and a
new octupole coupling involving the KK tensor field at graph (a3).}
\label{cubspin1g0loop}
\end{figure}

As can be seen in figure \ref{cubspin1g0loop} we have $10$ graphs of
one-graviton exchange in this sector, the majority of which already
involve time derivatives to be applied. Consistent with former works by
one of the authors we keep all of the higher order time derivative terms
that emerge in the evaluations of the graphs, and they will be treated
properly via redefinitions of the position and the rotational variables
as shown in \cite{Levi:2014sba}.
Notice that we have here the 4 graphs that appeared at the LO with the
quadratic time insertions on the propagators at graphs 1(a7)-(a10), and a
new octupole coupling involving the KK tensor field at graph 1(a3).

The graphs in figure \ref{cubspin1g0loop} are evaluated as follows:
\begin{align}
\text{Fig.~2(a1)}&=-C_{1(BS^3)}\frac{G}{r^4}\frac{m_2}{m_1^2} 
\Big[
 \vec{S}_1 \cdot\vec{v}_1 \times \vec{v}_2 \big(
 2\vec{S}_1 \cdot\vec{v}_2\ \vec{S}_1 \cdot \vec{n}
 + \vec{v}_2 \cdot \vec{n}\big(S_1^2 - 5\big(S_1 \cdot \vec{n}\big)^2\big) \nn &\\
&\qquad \qquad \qquad \qquad \qquad \qquad \qquad
\qquad \qquad \qquad \qquad \qquad \qquad 
- \vec{S}_1 \cdot \vec{v}_1\ \vec{S}_1 \cdot \vec{n}\big)\nn &\\
& +\vec{S}_1 \cdot\vec{v}_1 \times \vec{n}\Big(
\big(S^2_1 - 5\big(\vec{S}_1 \cdot\vec{n}\big)^2\big)  
\vec{v}_1 \cdot\vec{v}_2 + \vec{S}_1 \cdot \vec{v}_2 \big(\vec{S}_1 \cdot \vec{v}_2 - 5\vec{S}_1 \cdot \vec{n}\ \vec{v}_2 \cdot \vec{n}\big)\nn &\\
&- \vec{S}_1 \cdot \vec{v}_2 \big(\vec{S}_1 \cdot \vec{v}_1 - 5\vec{S}_1 \cdot\vec{n}\ \vec{v}_1 \cdot \vec{n}\big)\Big)
+\vec{S}_1 \cdot\vec{v}_2 \times \vec{n}
\Big(\frac{1}{2}S^2_1\big(v^2_1 + v^2_2\big)\nn &\\
& - \vec{S}_1 \cdot\vec{v}_1\big(\vec{S}_1 \cdot\vec{v}_2
- 5\vec{S}_1 \cdot\vec{n}\ \vec{v}_2 \cdot\vec{n}\big)
- \frac{5}{2}\big(\vec{S}_1 \cdot\vec{n}\big)^2\big(v^2_1 + v^2_2\big)\Big) \nn &\\
&-\frac{1}{2}\vec{v}_1 \cdot\vec{v}_2 \times\vec{n}\ \vec{S}_1 \cdot\vec{v}_1
\big(S^2_1 - 5\big(\vec{S}_1 \cdot\vec{n}\big)^2 \big)\Big] \nn &\\
&-\frac{1}{3}C_{1(BS^3)}\frac{G}{r^3}\frac{m_2}{m_1^2}\ \Big[
\vec{S}_1 \cdot \vec{v}_1 \times \vec{a}_2 \big(S_ 1^2 - 3\big(\vec{S}_1 \cdot \vec{n}\big)^2\big) - 3\vec{S}_1 \cdot \vec{v}_1 \times\vec{n} \vec{S}_1 \cdot \vec{a}_2\ \vec{S}_1 \cdot \vec{n} &\\
&+3\vec{S}_1 \cdot \vec{a}_2 \times\vec{n} \vec{S}_1 \cdot \vec{v}_1\ \vec{S}_1 \cdot \vec{n}\Big],&\\
\text{Fig.~2(a2)}&=\frac{1}{2}C_{1(BS^3)}\frac{G}{r^4}\frac{m_2}{m_1^2}\
\Big[\vec{S}_1\cdot \vec{v}_1\times \vec{n}\Big(
S_1^2\big(v_1^2+3v_2^2\big)
-2\vec{S}_1\cdot\vec{n}
\big(\vec{S}_1\cdot\vec{v}_2-5\ 
\vec{S}_1\cdot\vec{n}\ \vec{v}_2\cdot\vec{n}\big) \nn &\\
&
-5(\vec{S}_1\cdot\vec{n})^2\big(v_1^2+3v_2^2\big)\Big)
-2\vec{S}_1\cdot \vec{v}_1\times \vec{v}_2\ \vec{S}_1 \cdot\vec{v}_1\ \vec{S}_1 \cdot \vec{n}
\Big],&\\
\text{Fig.~2(a3)}&= -C_{1(BS^3)}\frac{G}{r^4}\frac{m_2}{m_1^2}
\Big[\big(\vec{S}_1\cdot \vec{v}_1\times \vec{n}\ v_2^2
-\vec{S}_1\cdot \vec{v}_2\times \vec{n}\  \vec{v}_1\cdot\vec{v}_2\big)
\big(S_1^2-5(\vec{S}_1\cdot\vec{n})^2\big) \Big]\nn &\\
&
+ C_{1(BS^3)}\frac{G}{r^3}\frac{m_2}{m_1^2}\Big[
\vec{S}_1\cdot \vec{v}_2\times \vec{n}\big(
\vec{S}_1\cdot\vec{v}_2 \dot{\vec{S}}_1\cdot\vec{n}
+ \dot{\vec{S}}_1\cdot\vec{v}_2 {\vec{S}}_1\cdot\vec{n} \big) \nn &\\
&
+ \dot{\vec{S}}_1\cdot \vec{v}_2\times \vec{n}
\ {\vec{S}}_1\cdot\vec{v}_2\ \vec{S}_1\cdot\vec{n}   \Big],&\\
\text{Fig.~2(a4)}&= \frac{3}{2}C_{1(ES^2)}\frac{G}{r^4}\frac{1}{m_1}
\Big[2\vec{S}_1 \cdot \vec{S}_2 \times \vec{v}_2 
\Big(\vec{S}_1 \cdot \vec{v}_1\ \vec{v}_1 \cdot \vec{n}
 -\vec{S}_1\cdot \vec{n}\big(3 v^2_1 + v^2_2\big)\ \Big)\nn&\\
&
+2\vec{S}_2 \cdot \vec{v}_1 \times \vec{v}_2
\Big(2 S^2_1 \ \vec{v}_1 \cdot \vec{n} 
- \vec{S}_1 \cdot \vec{v}_1\ \vec{S}_1 \cdot \vec{n} \Big)\nn&\\
&
- \vec{S}_2 \cdot \vec{v}_2 \times \vec{n}
\Big( S_1^2\big(5 v^2_1 + v^2_2 - 10
\big(\vec{v}_1 \cdot \vec{n} \big)^2\big)
- 2\vec{S}_1 \cdot \vec{v}_1 \big(\vec{S}_1 \cdot \vec{v}_1
- 5\vec{S}_1 \cdot \vec{n}\ \vec{v}_1 \cdot \vec{n} \big)
\nn &\\
& - 5\big(\vec{S}_1 \cdot \vec{n} 
\big)^2 \big(3 v^2_1 + v^2_2\big)\Big) \Big]
\nn &\\
&+C_{1(ES^2)}\frac{G}{r^3}\frac{1}{m_1} \Big[
2\ \vec{S}_1 \cdot \vec{S}_2 \times \vec{v}_2\
\big( \vec{S}_1 \cdot \vec{a}_1 +\dot{\vec{S}}_1 \cdot \vec{v}_1 \big)
+2\dot{\vec{S}}_1 \cdot \vec{S}_2 \times \vec{v}_2\ 
\vec{S}_1 \cdot \vec{v}_1  \nn &\\
& + 4\vec{S}_2 \cdot \vec{v}_1 \times \vec{v}_2\ 
\dot{\vec{S}}_1 \cdot \vec{S}_1\
+ 2\vec{S}_2 \cdot \vec{v}_2 \times \vec{a}_1\ S^2_1\
+ \vec{S}_2 \cdot \vec{v}_2 \times \vec{a}_2 \big( S^2_1
-3 \big( \vec{S}_1 \cdot \vec{n} \big)^2 \big) \nn&\\
&
-6\vec{S}_2 \cdot \vec{v}_2 \times \vec{n} 
\Big(S^2_1\ \vec{a}_1 \cdot \vec{n}
-2\dot{\vec{S}}_1 \cdot \vec{S}_1\ \vec{v}_1 \cdot \vec{n}
+ \vec{S}_1 \cdot \vec{v}_1\ \dot{\vec{S}}_1 \cdot \vec{n}
+ \dot{\vec{S}}_1 \cdot \vec{v}_1\ \vec{S}_1 \cdot \vec{n}  \nn &\\
&
+\vec{S}_1 \cdot \vec{a}_1\ \vec{S}_1 \cdot \vec{n}  \Big)\Big] \nn &\\
& -4C_{1(ES^2)}\frac{G}{r^2} \frac{1}{m_1}
\vec{S}_2 \cdot \vec{v}_2 \times \vec{n} \Big( \dot{S}^2_1
+ \ddot{\vec{S}}_1 \cdot \vec{S}_1\Big),&\\
\text{Fig.~2(a5)}&= -\frac{3}{2}C_{1(ES^2)}\frac{G}{r^4}\frac{1}{m_1} \Big[
2\vec{S}_1 \cdot \vec{S}_2 \times \vec{v}_1
\big(\vec{S}_1 \cdot \vec{v}_1\ \vec{v}_1 \cdot \vec{n}
- \vec{S}_1 \cdot \vec{n}\, v^2_1\big)
\nn &\\
& -6 \vec{S}_1 \cdot \vec{S}_2 \times \vec{v}_2\ 
\vec{S}_1 \cdot \vec{n}\ \vec{v}_1 \cdot \vec{v}_2
+\vec{S}_2 \cdot \vec{v}_1 \times \vec{v}_2\Big(
S^2_1 \vec{v}_2 \cdot \vec{n}
+ 2\vec{S}_1 \cdot \vec{v}_2\ \vec{S}_1 \cdot \vec{n}
\nn &\\
&
- 5\big(\vec{S}_1 \cdot \vec{n}\big)^2 \vec{v}_2 \cdot \vec{n}\Big)
-\vec{S}_2 \cdot \vec{v}_1 \times \vec{n}
\Big(S^2_1\ \big(3v^2_1-10 \big(\vec{v}_1 \cdot \vec{n}\big)^2\big)
- 2\big(\vec{S}_1 \cdot \vec{v}_1\big)^2
\nn &\\
&
+ 10\vec{S}_1 \cdot \vec{v}_1\ \vec{S}_1 \cdot \vec{n}\ 
\vec{v}_1 \cdot \vec{n}
- 5\big(\vec{S}_1 \cdot \vec{n}\big)^2\ \vec{v}^2_1 \Big) \nn&\\
&
-3 \vec{S}_2 \cdot \vec{v}_2 \times \vec{n}\
\Big( S^2_1 - 5\big(\vec{S}_1 \cdot \vec{n} \big)^2 \Big)  
\vec{v}_1 \cdot \vec{v}_2\
\Big] \nn&\\
&
-\frac{1}{2} C_{1(ES^2)}\frac{G}{r^3}\frac{1}{m_1}\Big[
\vec{S}_1 \cdot \vec{S}_2 \times \vec{v}_2
\big(4\ \dot{\vec{S}}_1 \cdot \vec{v}_2
- 3\dot{\vec{S}}_1 \cdot \vec{n}\ \vec{v}_2 \cdot \vec{n}\big)
\nn &\\
& +\dot{\vec{S}}_1 \cdot \vec{S}_2 \times \vec{v}_2 
\big(4\ \vec{S}_1 \cdot \vec{v}_2
- 3\vec{S}_1 \cdot \vec{n}\ \vec{v}_2 \cdot \vec{n}\big)
- 3\vec{S}_1 \cdot \vec{S}_2 \times \vec{n}
\Big(\vec{S}_1 \cdot \vec{v}_1\ \vec{a}_1 \cdot \vec{n}
\nn&\\
& +\dot{\vec{S}}_1 \cdot \vec{v}_1\ \vec{v}_1 \cdot \vec{n}
+ \vec{S}_1 \cdot \vec{a}_1\ \vec{v}_1 \cdot \vec{n}
- 2\vec{S}_1 \cdot \vec{n}\ \vec{v}_1 \cdot \vec{a}_1
-\ \dot{\vec{S}}_1 \cdot \vec{n} \,v^2_1 \Big)\nn &\\
& -3\dot{\vec{S}}_1 \cdot \vec{S}_2 \times \vec{n}
\big(\vec{S}_1 \cdot \vec{v}_1\ \vec{v}_1 \cdot \vec{n}
-\vec{S}_1 \cdot \vec{n} \, v^2_1 \big) \nn &\\
& +4\big(\vec{S}_2 \cdot \vec{v}_1 \times \vec{a}_2
+ \dot{\vec{S}}_2 \cdot \vec{v}_1 \times \vec{v}_2\big)
\Big(S^2_1 - 3\big(\vec{S}_1 \cdot \vec{n}\big)^2\Big)
\nn &\\
&-3 \vec{S}_2 \cdot \vec{v}_1 \times \vec{n} \Big(
2S^2_1\ \vec{a}_1 \cdot \vec{n}
+ 4\dot{\vec{S}}_1 \cdot \vec{S}_1\ \vec{v}_1 \cdot \vec{n}
- \vec{S}_1 \cdot \vec{v}_1 \ \dot{\vec{S}}_1 \cdot \vec{n}
- \dot{\vec{S}}_1 \cdot \vec{v}_1\ \vec{S}_1 \cdot \vec{n} \nn &\\
&
- \vec{S}_1 \cdot \vec{a}_1\ \vec{S}_1 \cdot \vec{n} \Big)
- 3\vec{S}_2 \cdot \vec{a}_1 \times \vec{n}
\big( 2S^2_1\ \vec{v}_1 \cdot \vec{n}
- \vec{S}_1 \cdot \vec{v}_1\ \vec{S}_1 \cdot \vec{n} \big) \nn&\\
& +3\vec{S}_2 \cdot \vec{v}_2 \times \vec{n} \big(
8\dot{\vec{S}}_1 \cdot \vec{S}_1\ \vec{v}_2 \cdot \vec{n}
-3 \vec{S}_1 \cdot \vec{v}_2 \dot{\vec{S}}_1 \cdot \vec{n}
-3 \dot{\vec{S}}_1 \cdot \vec{v}_2\ \vec{S}_1 \cdot \vec{n} \big)
\Big] \nn &\\
& +2C_{1(ES^2)}\frac{G}{r^2} \frac{1}{m_1} \Big[
\big( \vec{S}_1 \cdot \vec{S}_2 \times \vec{a}_2
+ \vec{S}_1 \cdot \dot{\vec{S}}_2 \times \vec{v}_2 \big) \dot{\vec{S}}_1 \cdot \vec{n}
\nn &\\
& +\big( \dot{\vec{S}}_1 \cdot \dot{\vec{S}}_2 \times \vec{v}_2
+ \dot{\vec{S}}_1 \cdot \vec{S}_2 \times \vec{a}_2 \big) \vec{S}_1 \cdot \vec{n}
- 2\big( \dot{\vec{S}}_2 \cdot \vec{v}_2 \times \vec{n}
+ \vec{S}_2 \cdot \vec{a}_2 \times \vec{n} \big)\dot{\vec{S}}_1 \cdot \vec{S}_1
\Big],&\\
\text{Fig.~2(a6)}&= -3C_{1(ES^2)}\frac{G}{r^4} \frac{1}{m_1} \Big[
2\vec{S}_1 \cdot\vec{S}_2 \times \vec{v}_1\
\vec{S}_1 \cdot \vec{n}\ \vec{v}_1 \cdot \vec{v}_2
-2\vec{S}_1 \cdot \vec{S}_2 \times \vec{v}_2\ \vec{S}_1 \cdot \vec{n} \,v_1^2\nn&\\
&+\Big(\vec{S}_2 \cdot \vec{v}_1 \times \vec{n}\ \vec{v}_1 \cdot \vec{v}_2
 - \vec{S}_2 \cdot \vec{v}_2 \times \vec{n}\ v^2_1 \Big)
 \Big(S^2_1 - 5\big(\vec{S}_1 \cdot \vec{n}\big)^2\Big) \Big]\nn&\\
& +C_{1(ES^2)}\frac{G}{m_1r^3} \Big[
\vec{S}_1 \cdot \vec{S}_2 \times \vec{v}_1 \dot{\vec{S}}_1 \cdot \vec{v}_2
+\dot{\vec{S}}_1 \cdot \vec{S}_2 \times \vec{v}_1 \vec{S}_1 \cdot \vec{v}_2
+2\vec{S}_1 \cdot \vec{S}_2 \times \vec{a}_1  \vec{S}_1 \cdot \vec{v}_2
 \nn &\\
& - 2 \vec{S}_1 \cdot \vec{S}_2 \times \vec{v}_2
\big(\dot{\vec{S}}_1 \cdot \vec{v}_1 + \vec{S}_1 \cdot \vec{a}_1 \big)
 -2\dot{\vec{S}}_1 \cdot \vec{S}_2 \times \vec{v}_2\ \vec{S}_1 \cdot \vec{v}_1 \nn &\\
& + 3\vec{S}_1 \cdot \vec{S}_2 \times \vec{n}
\Big( \vec{S}_1 \cdot \vec{n}\ \vec{a}_1 \cdot \vec{v}_2 - \vec{S}_1 \cdot \vec{v}_2\ \vec{a}_1 \cdot \vec{n} 
+ \dot{\vec{S}}_1 \cdot \vec{n}\ \vec{v}_1 \cdot \vec{v}_2 \Big) \nn &\\
&
+3 \dot{\vec{S}}_1 \cdot \vec{S}_2 \times \vec{n}\
\vec{S}_1 \cdot \vec{n}\ \vec{v}_1 \cdot \vec{v}_2
-6 \vec{S}_2 \cdot \vec{v}_1 \times \vec{v}_2 \dot{\vec{S}}_1 \cdot \vec{S}_1
 +2\vec{S}_2 \cdot \vec{v}_2 \times \vec{a}_1 S^2_1\nn&\\
& 
 +3 \vec{S}_2 \cdot \vec{v}_1 \times \vec{n}
 \Big(2\dot{\vec{S}}_1 \cdot \vec{S}_1\ \vec{v}_2 \cdot \vec{n}
 -\vec{S}_1 \cdot \vec{v}_2\ \dot{\vec{S}}_1 \cdot \vec{n}
  - \dot{\vec{S}}_1 \cdot \vec{v}_2\ \vec{S}_1 \cdot \vec{n} \Big)
 \nn&\\
 & 
-3\vec{S}_2 \cdot \vec{v}_2 \times \vec{n} \Big(S^2_1\ \vec{a}_1 \cdot \vec{n}
 +4\dot{\vec{S}}_1 \cdot \vec{S}_1\ \vec{v}_1 \cdot \vec{n}
\nn&\\
 &
-2 \vec{S}_1 \cdot \vec{v}_1 \ \dot{\vec{S}}_1 \cdot \vec{n}
- 2\dot{\vec{S}}_1 \cdot \vec{v}_1\ \vec{S}_1 \cdot \vec{n}
- 2\vec{S}_1 \cdot \vec{n}\ \vec{S}_1 \cdot \vec{a}_1\Big)\nn &\\
&+ 3\vec{S}_2 \cdot \vec{a}_1 \times \vec{n} \big(S^2_1 \vec{v}_2 \cdot \vec{n}
 - \vec{S}_1 \cdot \vec{v}_2 \ \vec{S}_1 \cdot \vec{n} \big)\Big]\nn&\\
&
-C_{1(ES^2)}\frac{G}{r^2} \frac{1}{m_1}\Big[
\vec{S}_1 \cdot \vec{S}_2 \times \vec{n}\ \ddot{\vec{S}}_1 \cdot \vec{v}_2
+2 \dot{\vec{S}}_1 \cdot \vec{S}_2 \times \vec{n}\ \dot{\vec{S}}_1 \cdot \vec{v}_2
+ \ddot{\vec{S}}_1 \cdot \vec{S}_2 \times \vec{n}\ \vec{S}_1 \cdot \vec{v}_2 \nn&\\
&-2 \vec{S}_2 \cdot \vec{v}_2 \times \vec{n}\
\big( \dot{S}^2_1+\ddot{\vec{S}}_1 \cdot \vec{S}_1   \big)
 \Big],&\\
\text{Fig.~2(a7)}&= \frac{1}{2}C_{1(BS^3)}\frac{G}{r^4} \frac{m_2}{m_1^2}
\Big[
\vec{S}_1\cdot \vec{v}_1\times \vec{v}_2 \Big({S}_1^2 \vec{v}_2 \cdot \vec{n}
+ 2\vec{S}_1 \cdot \vec{v}_2\ \vec{S}_1 \cdot \vec{n}
-5 \big( \vec{S}_1 \cdot \vec{n} \big)^2 \vec{v}_2 \cdot \vec{n}\Big) \nn&\\
&-\vec{S}_1\cdot \vec{v}_2\times \vec{n}
\Big( S_1^2\big(\vec{v}_1 \cdot \vec{v}_2-5\vec{v}_1 \cdot \vec{n}\ \vec{v}_2 \cdot \vec{n}\big)
 + 2 \vec{S}_1 \cdot \vec{v}_1\ \vec{S}_1 \cdot \vec{v}_2 \nn&\\
&-10\vec{S}_1 \cdot \vec{v}_1\ \vec{S}_1 \cdot \vec{n}\ \vec{v}_2 \cdot \vec{n}
-10 \vec{S}_1 \cdot \vec{v}_2\ \vec{S}_1 \cdot \vec{n}\ \vec{v}_1 \cdot \vec{n} \nn&\\
&- 5\big(\vec{S}_1 \cdot \vec{n} \big)^2 \big(\vec{v}_1 \cdot \vec{v}_2
-7\vec{v}_1 \cdot \vec{n}\ \vec{v}_2 \cdot \vec{n}  \big)\Big)\Big] \nn &\\
&
+\frac{1}{6}C_{1(BS^3)}\frac{G}{r^3}\frac{m_2}{m_1^2}
\Big[ \vec{S}_1\cdot \vec{v}_1\times \vec{a}_2
\big({S}_1^2-3 \big(\vec{S}_1 \cdot \vec{n}\big)^2 \big)\nn &\\
&-6\vec{S}_1\cdot \vec{v}_2\times \vec{n}
\big( \dot{\vec{S}}_1\cdot \vec{S}_1 \vec{v}_2 \cdot \vec{n}
+\dot{\vec{S}}_1\cdot \vec{v}_2 \vec{S}_1\cdot \vec{n}
+\vec{S}_1\cdot \vec{v}_2 \dot{\vec{S}}_1\cdot \vec{n}
-5\dot{\vec{S}}_1\cdot \vec{n}\ \vec{S}_1\cdot \vec{n}\ \vec{v}_2\cdot \vec{n} \big) \nn&\\
&
+ 3\vec{S}_1\cdot \vec{a}_2\times \vec{n}
\big( {S}_1^2\vec{v}_1\cdot\vec{n}+2 \vec{S}_1 \cdot \vec{v}_1\ \vec{S}_1 \cdot \vec{n}
-5\big(\vec{S}_1 \cdot \vec{n}\big)^2 \vec{v}_1 \cdot \vec{n} \big)\nn &\\
&-3 \dot{\vec{S}}_1\cdot \vec{v}_2\times \vec{n}
\big( {S}_1^2 \vec{v}_2\cdot\vec{n}+2 \vec{S}_1 \cdot \vec{v}_2\ \vec{S}_1 \cdot \vec{n}
-5\big(\vec{S}_1 \cdot \vec{n}\big)^2 \vec{v}_2\cdot \vec{n} \big) \Big]\nn &\\
&
-\frac{1}{6}C_{1(BS^3)}\frac{G}{r^2}\frac{m_2}{m_1^2}
\Big[2\vec{S}_1\cdot \vec{a}_2\times \vec{n}
\big(\dot{\vec{S}}_1\cdot \vec{S}_1-3 \dot{\vec{S}}_1 \cdot \vec{n}\ \vec{S}_1 \cdot \vec{n} \big) \nn&\\
&+ \dot{\vec{S}}_1\cdot \vec{a}_2\times \vec{n}\big({S}_1^2-3 (\vec{S}_1 \cdot \vec{n})^2 \big) \Big]
,&\\
\text{Fig.~2(a8)}&= \frac{1}{2}C_{1(BS^3)}\frac{G}{r^4} \frac{m_2}{m_1^2} \Big[
\vec{S}_1 \cdot \vec{v}_1 \times \vec{v}_2
\Big( {S}_1^2\vec{v}_1 \cdot \vec{n} + 2\vec{S}_1 \cdot \vec{v}_1\ \vec{S}_1 \cdot \vec{n}\
-5(\vec{S}_1 \cdot \vec{n})^2 \vec{v}_1 \cdot \vec{n}\Big) \nn&\\
& +\vec{S}_1 \cdot \vec{v}_1 \times \vec{n}
\Big(S_1^2 \big(\vec{v}_1 \cdot \vec{v}_2-5\ \vec{v}_1 \cdot \vec{n}\ \vec{v}_2 \cdot \vec{n} \big)
+ 2 \vec{S}_1 \cdot \vec{v}_1\ \vec{S}_1 \cdot \vec{v}_2 \nn&\\
& -10\vec{S}_1 \cdot \vec{v}_1\ \vec{S}_1 \cdot \vec{n}\ \vec{v}_2 \cdot \vec{n}
-10 \vec{S}_1 \cdot \vec{v}_2\ \vec{S}_1 \cdot \vec{n}\ \vec{v}_1 \cdot \vec{n}   \nn&\\
& - 5( \vec{S}_1 \cdot \vec{n})^2 \big( \vec{v}_1 \cdot \vec{v}_2
- 7  \vec{v}_1 \cdot \vec{n}\ \vec{v}_2 \cdot \vec{n} \big)\Big) \Big]\nn &\\
&
- \frac{1}{6}C_{1(BS^3)}\frac{G}{r^3} \frac{m_2}{m_1^2} \Big[
2\vec{S}_1\cdot \vec{v}_1\times \vec{v}_2
\big( \dot{\vec{S}}_1 \cdot \vec{S}_1 - 3\dot{\vec{S}}_1 \cdot \vec{n}\ \vec{S}_1 \cdot \vec{n} \big)\nn&\\
&
- \big(\vec{S}_1\cdot \vec{v}_2\times \vec{a}_1 - \dot{\vec{S}}_1\cdot \vec{v}_1\times \vec{v}_2\big)
\big({S}_1^2 - 3\big(\vec{S}_1 \cdot \vec{n}\big)^2 \big)
-6 \vec{S}_1\cdot \vec{v}_1\times \vec{n}\big( \dot{\vec{S}}_1\cdot \vec{S}_1\ \vec{v}_2 \cdot \vec{n}\nn &\\
&
+ \dot{\vec{S}}_1\cdot \vec{v}_2\ \vec{S}_1\cdot \vec{n}
+\vec{S}_1\cdot \vec{v}_2\ \dot{\vec{S}}_1\cdot \vec{n}
-5\dot{\vec{S}}_1\cdot \vec{n}\ \vec{S}_1\cdot \vec{n}\ \vec{v}_2\cdot \vec{n} \big)\nn&\\
& - 3 \big(\vec{S}_1\cdot \vec{a}_1\times \vec{n} + \dot{\vec{S}}_1\cdot \vec{v}_1\times \vec{n} \big)
\big( {S}_1^2\ \vec{v}_2 \cdot \vec{n}+2\vec{S}_1 \cdot \vec{v}_2\ \vec{S}_1 \cdot \vec{n}
-5(\vec{S}_1 \cdot \vec{n})^2\vec{v}_2 \cdot \vec{n} \big)\Big]
,&\\
\text{Fig.~2(a9)}&=
-\frac{3}{2}C_{1(ES^2)}\frac{G}{r^4}\frac{1}{m_1} \Big[
2 \vec{S}_1 \cdot \vec{S}_2 \times \vec{v}_2 \Big(
\vec{S}_1 \cdot \vec{v}_1\ \vec{v}_2 \cdot \vec{n}\
+\vec{S}_1 \cdot \vec{v}_2\ \vec{v}_1 \cdot \vec{n}\
\nn&\\
&+ \vec{S}_1 \cdot \vec{n}\ \vec{v}_1 \cdot \vec{v}_2\
- 5\vec{S}_1 \cdot \vec{n}\ \vec{v}_1 \cdot \vec{n}\ \vec{v}_2 \cdot \vec{n}\ \Big)
- \vec{S}_2 \cdot \vec{v}_1 \times \vec{v}_2 \Big( S^2_1 \vec{v}_2 \cdot \vec{n}
+2 \vec{S}_1 \cdot \vec{v}_2\ \vec{S}_1 \cdot \vec{n}\ \nn&\\
&
-5\big( \vec{S}_1 \cdot \vec{n} \big)^2 \vec{v}_2 \cdot \vec{n}\Big)
- \vec{S}_2 \cdot \vec{v}_2 \times \vec{n}\
\Big(S_1^2 \big(\vec{v}_1 \cdot \vec{v}_2\
- 5\ \vec{v}_1 \cdot \vec{n}\  \vec{v}_2 \cdot \vec{n}\big)
-2\vec{S}_1 \cdot \vec{v}_1\ \vec{S}_1 \cdot \vec{v}_2
\nn&\\
&
+ 10 \vec{S}_1 \cdot \vec{v}_1\ \vec{S}_1 \cdot \vec{n}\ \vec{v}_2 \cdot \vec{n}
 + 10 \vec{S}_1 \cdot \vec{v}_2\ \vec{S}_1 \cdot \vec{n}\ \vec{v}_1 \cdot \vec{n} \Big)\nn&\\
 &
 + 5\big(\vec{S}_1 \cdot \vec{n} \big)^2 \big(\vec{v}_1 \cdot \vec{v}_2\
 - 7\ \vec{v}_1 \cdot \vec{n}\  \vec{v}_2 \cdot \vec{n}  \big)\Big)\Big]
 \nn&\\
&
+ \frac{1}{2} C_{1(ES^2)}\frac{G}{r^3} \frac{1}{m_1}\Big[
2 \vec{S}_1 \cdot \vec{S}_2 \times \vec{v}_2
\big( \dot{\vec{S}}_1 \cdot \vec{v}_2
- 3\dot{\vec{S}}_1 \cdot \vec{n}\ \vec{v}_2 \cdot \vec{n} \big)\nn&\\
& -2 \Big(\vec{S}_1 \cdot \vec{S}_2 \times \vec{a}_2
+ \vec{S}_1 \cdot \dot{\vec{S}}_2 \times \vec{v}_2 \Big)
\Big(\vec{S}_1 \cdot \vec{v}_1 - 3\ \vec{S}_1 \cdot \vec{n}\ \vec{v}_1 \cdot \vec{n} \Big)
\nn &\\
& +2 \dot{\vec{S}}_1 \cdot \vec{S}_2 \times \vec{v}_2
\big( \vec{S}_1 \cdot \vec{v}_2 -3\vec{S}_1 \cdot \vec{n}\ \vec{v}_2 \cdot \vec{n} \big) \nn&\\
&-\Big(\vec{S}_2 \cdot \vec{v}_1 \times \vec{a}_2
+ \dot{\vec{S}}_2 \cdot \vec{v}_1 \times \vec{v}_2\Big)
\Big(S^2_1 + 3 \big( \vec{S}_1 \cdot \vec{n} \big)^2 \Big)
+6\vec{S}_2 \cdot \vec{v}_2 \times \vec{n}
\Big( \dot{\vec{S}}_1 \cdot \vec{S}_1\ \vec{v}_2 \cdot \vec{n}
 \nn &\\
& - \dot{\vec{S}}_1 \cdot \vec{v}_2\ \vec{S}_1 \cdot \vec{n}
- \vec{S}_1 \cdot \vec{v}_2\ \dot{\vec{S}}_1 \cdot \vec{n}
+ 5\dot{\vec{S}}_1 \cdot \vec{n}\ \vec{S}_1 \cdot \vec{n}\ \vec{v}_2 \cdot \vec{n} \Big) \nn &\\
& -3 \Big(\dot{\vec{S}}_2 \cdot \vec{v}_2 \times \vec{n}
+ \vec{S}_2 \cdot \vec{a}_2 \times \vec{n} \Big)
\Big( S^2_1\ \vec{v}_1 \cdot \vec{n} - 2\vec{S}_1 \cdot \vec{v}_1\ \vec{S}_1 \cdot \vec{n}
+ 5 \big(\vec{S}_1 \cdot \vec{n} \big)^2\ \vec{v}_1 \cdot \vec{n} \Big) \Big] \nn &\\
& -C_{1(ES^2)}\frac{G}{r^2}\frac{1}{m_1}\Big[
\vec{S}_1 \cdot \vec{S}_2 \times \vec{a}_2\ \dot{\vec{S}}_1 \cdot \vec{n}
+ \vec{S}_1 \cdot \dot{\vec{S}}_2 \times \vec{v}_2\ \dot{\vec{S}}_1 \cdot \vec{n}
+ \dot{\vec{S}}_1 \cdot \vec{S}_2 \times \vec{a}_2\ \vec{S}_1 \cdot \vec{n}\nn&\\
&
+\dot{\vec{S}}_1 \cdot \dot{\vec{S}}_2 \times \vec{v}_2\ \vec{S}_1 \cdot \vec{n}
-\Big( \dot{\vec{S}}_2 \cdot \vec{v}_2 \times \vec{n}
+ \vec{S}_2 \cdot \vec{a}_2 \times \vec{n} \Big)
\Big( \dot{\vec{S}}_1 \cdot \vec{S}_1
+ 3\dot{\vec{S}}_1 \cdot \vec{n}\ \vec{S}_1 \cdot \vec{n} \Big) \Big]
,&\\
\text{Fig.~2(a10)}&=\frac{3}{2}C_{1(ES^2)}\frac{G}{r^4} \frac{1}{m_1}\Big[
2\vec{S}_1 \cdot \vec{S}_2 \times \vec{v}_1 \Big(
\vec{S}_1 \cdot\vec{v}_1\ \vec{v}_2 \cdot\vec{n}
+\vec{S}_1 \cdot\vec{v}_2\ \vec{v}_1 \cdot\vec{n}
+ \vec{S}_1 \cdot\vec{n} \vec{v}_1 \cdot\vec{v}_2\ \nn&\\
&
- 5\vec{S}_1 \cdot\vec{n}\ \vec{v}_1 \cdot\vec{n}\ \vec{v}_2 \cdot\vec{n} \Big)
- \vec{S}_2 \cdot \vec{v}_1 \times \vec{v}_2 \Big( S^2_1 \vec{v}_1 \cdot \vec{n}
- 2\vec{S}_1 \cdot \vec{v}_1\ \vec{S}_1 \cdot \vec{n} \nn &\\
&
+ 5\big( \vec{S}_1 \cdot \vec{n} \big)^2 \vec{v}_1 \cdot \vec{n} \Big)
-\vec{S}_2 \cdot \vec{v}_1 \times \vec{n} \Big(S_1^2\big(\vec{v}_1\cdot\vec{v}_2-
5\ \vec{v}_1 \cdot \vec{n}\ \vec{v}_2\cdot\vec{n} \big)
\nn&\\
&
- 2\vec{S}_1 \cdot \vec{v}_1\ \vec{S}_1 \cdot \vec{v}_2
+10 \vec{S}_1 \cdot\vec{v}_1\ \vec{S}_1 \cdot\vec{n} \vec{v}_2 \cdot\vec{n}
 +10 \vec{S}_1 \cdot\vec{v}_2\ \vec{S}_1 \cdot\vec{n}\vec{v}_1 \cdot\vec{n}
\nn&\\
&
+5\big(\vec{S}_1 \cdot \vec{n}\big)^2 \big(\vec{v}_1\cdot\vec{v}_2
- 7 \vec{v}_1 \cdot \vec{n}\ \vec{v}_2\cdot\vec{n}\big)\Big) \Big]
\nn&\\
&-\frac{1}{2} C_{1(ES^2)}\frac{G}{r^3}\frac{1}{m_1}\Big[
\vec{S}_1 \cdot\vec{S}_2\times\vec{v}_1
\big(\dot{\vec{S}}_1 \cdot\vec{v}_2 -3\dot{\vec{S}}_1 \cdot\vec{n}\ \vec{v}_2\cdot\vec{n}\big)
\nn&\\
&
+ 2\vec{S}_1 \cdot\vec{S}_2\times\vec{a}_1
\big(\vec{S}_1 \cdot\vec{v}_2 -3\vec{S}_1 \cdot\vec{n}\ \vec{v}_2\cdot\vec{n}\big)
+\dot{\vec{S}}_1 \cdot\vec{S}_2\times\vec{v}_1
\big(\vec{S}_1 \cdot\vec{v}_2 -3\vec{S}_1 \cdot\vec{n}\ \vec{v}_2\cdot\vec{n}\big)\nn&\\
&
- 2\vec{S}_1 \cdot\dot{\vec{S}}_2\times\vec{v}_1
\Big(\vec{S}_1 \cdot\vec{v}_1 -3\vec{S}_1 \cdot\vec{n}\ \vec{v}_1\cdot\vec{n}\Big)
- \vec{S}_1 \cdot\vec{S}_2\times\vec{v}_2
\big(\dot{\vec{S}}_1 \cdot\vec{v}_1 - 3\dot{\vec{S}}_1 \cdot\vec{n}\ \vec{v}_1\cdot\vec{n}\big)
\nn&\\
&
-\dot{\vec{S}}_1 \cdot\vec{S}_2\times\vec{v}_2
\Big(\vec{S}_1 \cdot\vec{v}_1 -3\vec{S}_1 \cdot\vec{n}\ \vec{v}_1\cdot\vec{n}\Big)
+3\vec{S}_1 \cdot \vec{S}_2 \times \vec{n}
\Big(\dot{\vec{S}}_1 \cdot\vec{v}_1\ \vec{v}_2 \cdot \vec{n}
\nn&\\
&
+\dot{\vec{S}}_1 \cdot\vec{v}_2\ \vec{v}_1 \cdot \vec{n}
+ \dot{\vec{S}}_1 \cdot\vec{n}\ \vec{v}_1 \cdot \vec{v}_2
- 5\dot{\vec{S}}_1\cdot\vec{n}\ \vec{v}_1\cdot\vec{n}\ \vec{v}_2\cdot\vec{n}\Big) \nn&\\
&+3\dot{\vec{S}}_1 \cdot \vec{S}_2 \times \vec{n}
\Big(\vec{S}_1 \cdot\vec{v}_1\ \vec{v}_2 \cdot \vec{n}
+ \vec{S}_1 \cdot\vec{v}_2\ \vec{v}_1 \cdot \vec{n}
+  \vec{S}_1 \cdot\vec{n}\ \vec{v}_1 \cdot\vec{v}_2 \nn&\\
&
- 5\vec{S}_1\cdot\vec{n}\ \vec{v}_1\cdot\vec{n}\ \vec{v}_2\cdot\vec{n}\Big)
- 2\vec{S}_2 \cdot\vec{v}_1\times\vec{v}_2
\big(\dot{\vec{S}}_1 \cdot\vec{S}_1 + 3\dot{\vec{S}}_1 \cdot\vec{n}\ \vec{S}_1 \cdot\vec{n}\big)
\nn&\\
&
- \vec{S}_2 \cdot\vec{a}_1\times\vec{v}_2\Big(S^2_1 + 3\big(\vec{S}_1 \cdot\vec{n}\big)^2 \Big)
+6\vec{S}_2 \cdot\vec{v}_1\times\vec{n}\Big(
\dot{\vec{S}}_1 \cdot\vec{S}_1\ \vec{v}_2 \cdot\vec{n}
- \vec{S}_1 \cdot\vec{v}_2\dot{\vec{S}}_1 \cdot\vec{n}\  \nn&\\
&
- \dot{\vec{S}}_1 \cdot\vec{v}_2\ \vec{S}_1 \cdot\vec{n}
+ 5\dot{\vec{S}}_1 \cdot\vec{n}\ \vec{S}_1 \cdot\vec{n}\ \vec{v}_2 \cdot\vec{n}\Big)
+3\vec{S}_2 \cdot\vec{a}_1\times\vec{n}\Big(S^2_1\vec{v}_2 \cdot\vec{n}
- 2\vec{S}_1\cdot\vec{v}_2\ \vec{S}_1\cdot\vec{n}
\nn&\\
&
+ 5\big(\vec{S}_1\cdot\vec{n}\big)^2 \vec{v}_2 \cdot\vec{n} \Big)
- 3\dot{\vec{S}}_2 \cdot\vec{v}_1\times\vec{n}
\big(S^2_1\ \vec{v}_1 \cdot\vec{n} - 2\vec{S}_1 \cdot\vec{v}_1\ \vec{S}_1 \cdot\vec{n}
+ 5(\vec{S}_1 \cdot\vec{n})^2 \vec{v}_1 \cdot\vec{n} \big)
\Big] \nn &\\
& +\frac{1}{2}C_{1(ES^2)}\frac{G}{r^2} \frac{1}{m_1}\Big[
\vec{S}_1 \cdot \dot{\vec{S}}_2 \times \vec{v}_1\ \dot{\vec{S}}_1 \cdot \vec{n}
+\dot{\vec{S}}_1 \cdot \dot{\vec{S}}_2 \times \vec{v}_1\ \vec{S}_1 \cdot \vec{n}\nn&\\
&
+ \vec{S}_1 \cdot \vec{S}_2 \times \vec{v}_2\ \ddot{\vec{S}}_1 \cdot \vec{n}
+ 2 \dot{\vec{S}}_1 \cdot \vec{S}_2 \times \vec{v}_2\ \dot{\vec{S}}_1 \cdot \vec{n}
+ \ddot{\vec{S}}_1 \cdot\vec{S}_2 \times\vec{v}_2\ \vec{S}_1 \cdot\vec{n} \nn&\\
&+\vec{S}_1 \cdot \vec{S}_2 \times \vec{n}
\big( \ddot{\vec{S}}_1 \cdot\vec{v}_2 - 3\ddot{\vec{S}}_1 \cdot\vec{n}\ \vec{v}_2 \cdot\vec{n}\big)
- \vec{S}_1 \cdot  \dot{\vec{S}}_2 \times \vec{n}
\Big( \dot{\vec{S}}_1 \cdot \vec{v}_1 - 3\dot{\vec{S}}_1 \cdot \vec{n}\ \vec{v}_1 \cdot \vec{n} \Big)
\nn&\\
&
+ 2\dot{\vec{S}}_1 \cdot \vec{S}_2 \times \vec{n}
\big( \dot{\vec{S}}_1 \cdot\vec{v}_2 - 3\dot{\vec{S}}_1 \cdot\vec{n}\ \vec{v}_2 \cdot\vec{n}\big)
- \dot{\vec{S}}_1 \cdot  \dot{\vec{S}}_2 \times \vec{n}
\Big( {\vec{S}}_1 \cdot \vec{v}_1 - 3{\vec{S}}_1 \cdot \vec{n}\ \vec{v}_1 \cdot \vec{n} \Big)\nn&\\
&
+ \ddot{\vec{S}}_1 \cdot \vec{S}_2 \times \vec{n}
\big(\vec{S}_1 \cdot\vec{v}_2 - 3\vec{S}_1 \cdot\vec{n}\ \vec{v}_2 \cdot\vec{n}\big)
+ 2\vec{S}_1 \cdot \dot{\vec{S}}_2 \times \vec{a}_1\ \vec{S}_1 \cdot \vec{n} \nn&\\
&
-2 \ \dot{\vec{S}}_2 \cdot \vec{v}_1 \times \vec{n}
\Big( \dot{\vec{S}}_1 \cdot \vec{S}_1 + 3\dot{\vec{S}}_1 \cdot \vec{n}\ \vec{S}_1 \cdot \vec{n} \Big)
- \dot{\vec{S}}_2 \cdot \vec{a}_1 \times \vec{n}
\Big( S^2_1 + 3\big(\vec{S}_1 \cdot \vec{n} \big)^2 \Big)\Big]
\nn&\\
&
-\frac{1}{2}C_{1(ES^2)}\frac{G}{r} \frac{1}{m_1}\Big[
\vec{S}_1 \cdot  \dot{\vec{S}}_2 \times \vec{n}\ \ddot{\vec{S}}_1 \cdot \vec{n}
+ 2\dot{\vec{S}}_1 \cdot \dot{\vec{S}}_2 \times \vec{n}\ \dot{\vec{S}}_1 \cdot \vec{n}
+ \ddot{\vec{S}}_1 \cdot  \dot{\vec{S}}_2 \times \vec{n}\ \vec{S}_1 \cdot \vec{n} \Big]
.&
\end{align}

Note that almost all these graphs
contain higher order time derivatives terms, notably
second order time derivatives, where graph 1(a10) even contains
third order ones.
Notice also that the value of graph 1(a5) will have to be supplemented 
with a piece that contains time derivatives of the spin, that appeared already 
in graph 2(a) of the LO in \cite{Levi:2014gsa}, but eventually did not 
contribute at the LO. At this order, as we will see here in section
\ref{finalaction} these terms actually contribute.

\subsection{Two-graviton exchange}

\begin{figure}[t]
\centering
\includegraphics[width=\textwidth]{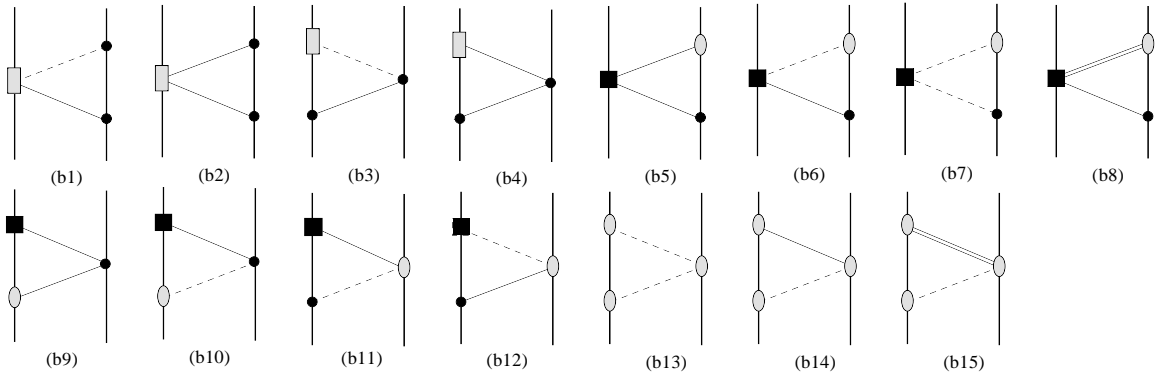}
\caption{The Feynman graphs of two-graviton exchange, which contribute 
to the NLO cubic-in-spin interaction at the 4.5PN order for 
maximally-rotating compact objects. 
The graphs should be included together with their mirror images, 
i.e.~with the worldline labels $1\leftrightarrow2$ exchanged.
These graphs include all relevant interactions among the spin-induced
quadrupole, octupole, and the mass and spin, in particular here at the
nonlinear level there are also interactions involving the various
multipoles on two different points of the worldline, which add up to
interactions that are cubic in the spin, such as a spin dipole and a
spin-induced quadrupole or two spin dipoles, on the same worldline,
which can already be seen as of the NLO spin-squared sector
\cite{Levi:2015msa,Levi:2015ixa}.
Consequently notice that there are nonlinearities originating from
gravitons sourced strictly from minimal coupling to the worldline as
shown in graphs (b13)-(b15). We also have here two new
two-graviton octupole couplings in graphs (b1), (b2).}
\label{cubspin2g0loop}
\end{figure}

As can be seen in figure \ref{cubspin2g0loop} we have $15$ graphs of
two-graviton exchange in this sector. Here the majority of the graphs do
not involve time derivatives. We have here two new two-graviton octupole
couplings in graphs 1(b1), 1(b2), and on the other hand we have here
nonlinearities originating from gravitons sourced strictly from minimal
coupling to the worldline as in graphs 1(b13)-1(b15).

The graphs in figure \ref{cubspin2g0loop} are evaluated as follows:
\begin{align}
\text{Fig.~3(b1)}&=C_{1(BS^3)}\frac{G^2}{r^5}\frac{m_2^2}{m_1^2}
\,\vec{S}_1\cdot \vec{v}_2\times \vec{n}\ \Big[9S_1^2-50(\vec{S}_1\cdot\vec{n})^2\Big]  ,&\\
\text{Fig.~3(b2)}&=-\frac{1}{3}C_{1(BS^3)}\frac{G^2}{r^5}\frac{m_2^2}{m_1^2}
\,\vec{S}_1\cdot \vec{v}_1\times \vec{n}\ \Big[11S_1^2-54(\vec{S}_1\cdot\vec{n})^2\Big] ,&\\
\text{Fig.~3(b3)}&= C_{1(BS^3)}\frac{G^2}{r^5}\frac{m_2}{m_1}
\,\vec{S}_1\cdot \vec{v}_2\times \vec{n}\left[S_1^2-5(\vec{S}_1\cdot\vec{n})^2\right],&\\
\text{Fig.~3(b4)}&= -C_{1(BS^3)}\frac{G^2}{r^5}\frac{m_2}{m_1}
\,\vec{S}_1\cdot \vec{v}_1\times \vec{n}\left[S_1^2-5(\vec{S}_1\cdot\vec{n})^2\right],&\\
\text{Fig.~3(b5)}&= 8 C_{1(ES^2)}\frac{G^2}{r^5}\frac{m_2}{m_1}
\,\left[3\,\vec{S}_1\cdot \vec{S}_2\times \vec{v}_2 \,\vec{S}_1\cdot\vec{n}
+\vec{S}_2\cdot \vec{v}_2\times\vec{n}
\left[2S_1^2-9(\vec{S}_1\cdot\vec{n})^2\right]\right],&\\
\text{Fig.~3(b6)}&= C_{1(ES^2)}\frac{G^2}{r^5}\frac{m_2}{m_1}
\left[-23\vec{S}_1\cdot \vec{S}_2\times \vec{v}_1\,\vec{S}_1\cdot\vec{n}
+13\vec{S}_1\cdot \vec{S}_2\times \vec{v}_2\,\vec{S}_1\cdot\vec{n}\right.\nn&\\
&-\vec{S}_2\cdot \vec{v}_1\times \vec{n}\big(31S_1^2 - 66\big(\vec{S}_1 \cdot\vec{n}\big)^2\big)-
\vec{S}_1\cdot\vec{S}_2\times\vec{n}\left(10\vec{S}_1\cdot\vec{v}_1-51\vec{S}_1\cdot\vec{n}\,\vec{v}_1\cdot\vec{n}\right)\nn&\\
&+\left.\vec{S}_1\cdot\vec{S}_2\times\vec{n}\left(11\vec{S}_1\cdot\vec{v}_2-54\vec{S}_1\cdot\vec{n}\,\vec{v}_2\cdot\vec{n}\right)\right]&\nn\\
&-13C_{1(ES^2)}\frac{G^2}{r^4}\frac{m_2}{m_1}
\,\left[\vec{S}_1\cdot \dot{\vec{S}}_2\times \vec{n}\ \vec{S}_1\cdot\vec{n}\right],&\\
\text{Fig.~3(b7)}&= 2 C_{1(ES^2)}\frac{G^2}{r^5}\frac{m_2}{m_1}
\left[2\,\vec{S}_1\cdot \vec{S}_2\times \vec{v}_2\,\vec{S}_1\cdot\vec{n}
+\vec{S}_1\cdot \vec{S}_2\times \vec{n}
\left(\vec{S}_1\cdot\vec{v}_2-3\vec{S}_1\cdot\vec{n}\,\vec{v}_2\cdot\vec{n}\right)\right.\nn&\\
&\left.+\,\vec{S}_2\cdot \vec{v}_2\times \vec{n}\,
\Big(2 {S}_1^2-3(\vec{S}_1\cdot\vec{n})^2\Big)\right],&\\
\text{Fig.~3(b8)}&= -C_{1(ES^2)}\frac{G^2}{r^5}\frac{m_2}{m_1}
\left[2\,\vec{S}_1\cdot \vec{S}_2\times \vec{v}_2\,\vec{S}_1\cdot\vec{n}
+3\vec{S}_1\cdot \vec{S}_2\times \vec{n}
\left(\vec{S}_1\cdot\vec{v}_2-2\vec{S}_1\cdot\vec{n}\,\vec{v}_2\cdot\vec{n}\right)\right.\nn&\\
&\left.+\,\vec{S}_2\cdot \vec{v}_2\times \vec{n}
\left(5S_1^2-12(\vec{S}_1\cdot\vec{n})^2\right)\right],&\\
\text{Fig.~3(b9)}&= -C_{1(ES^2)}\frac{G^2}{r^5}\frac{m_2}{m_1}
\,\vec{S}_1\cdot \vec{v}_1\times \vec{n}\left[S_1^2-3(\vec{S}_1\cdot\vec{n})^2\right],&\\
\text{Fig.~3(b10)}&= C_{1(ES^2)}\frac{G^2}{r^5}\frac{m_2}{m_1}
\,\vec{S}_1\cdot \vec{v}_2\times \vec{n}\left[S_1^2-3(\vec{S}_1\cdot\vec{n})^2\right],&\\
\text{Fig.~3(b11)}&= -4C_{1(ES^2)}\frac{G^2}{r^5}
\,\vec{S}_2\cdot \vec{v}_1\times \vec{n}\left[S_1^2-3(\vec{S}_1\cdot\vec{n})^2\right],&\\
\text{Fig.~3(b12)}&= -12C_{1(ES^2)}\frac{G^2}{r^5}
\left[2\,\vec{S}_1\cdot \vec{S}_2\times \vec{v}_1\,\vec{S}_1\cdot\vec{n}
+\vec{S}_2\cdot \vec{v}_1\times \vec{n}\left(S_1^2-5(\vec{S}_1\cdot\vec{n})^2\right)\right]\nn&\\
&+12C_{1(ES^2)}\frac{G^2}{r^4}
\,\left[\vec{S}_1\cdot \vec{S}_2\times \vec{n} \,\dot{\vec{S}}_1\cdot\vec{n}
+ \dot{\vec{S}}_1\cdot \vec{S}_2\times \vec{n} \,\vec{S}_1\cdot\vec{n}\right],&\\
\text{Fig.~3(b13)}&= 2\frac{G^2}{r^5}
\,\left[\vec{S}_1\cdot \vec{v}_2\times \vec{n}
\left(\vec{S}_1\cdot\vec{S}_2-3\vec{S}_1\cdot\vec{n}\vec{S}_2\cdot\vec{n}\right)-\vec{S}_1 \cdot \vec{S}_2\ \vec{S}_1\cdot \vec{v}_1\times \vec{n}\right]\nn&\\
&+2\frac{G^2}{r^4}\left[\dot{\vec{S}}_1 \cdot \vec{n}\ \vec{S}_1 \cdot \vec{S}_2 \times\vec{n} - \vec{S}_1 \cdot \vec{n}\ \dot{\vec{S}}_1 \cdot \vec{S}_2 \times\vec{n} - \dot{\vec{S}}_1 \cdot \vec{S}_1 \times\vec{S}_2\right],&\\
\text{Fig.~3(b14)}&= -8\frac{G^2}{r^5}
\,\vec{S}_1\cdot \vec{v}_1\times \vec{n}
\left[\vec{S}_1\cdot\vec{S}_2-3\vec{S}_1\cdot\vec{n}\vec{S}_2\cdot\vec{n}\right],&\\
\text{Fig.~3(b15)}&= -\frac{G^2}{r^5}
\left[2\,\vec{S}_1\cdot \vec{S}_2\times \vec{n}\,\vec{S}_1\cdot\vec{v}_1
-\vec{S}_1\cdot \vec{v}_1\times \vec{n}
\left(5\,\vec{S}_1\cdot\vec{S}_2-9\vec{S}_1\cdot\vec{n}\,\vec{S}_2\cdot\vec{n}\right)\right.\nn&\\
&\left.+3\,\vec{S}_2\cdot \vec{v}_1\times \vec{n}\left(S_1^2-(\vec{S}_1\cdot\vec{n})^2\right)\right].&
\end{align}

\subsection{Cubic self-interaction}

\begin{figure}[t]
\centering
\includegraphics[width=\textwidth]{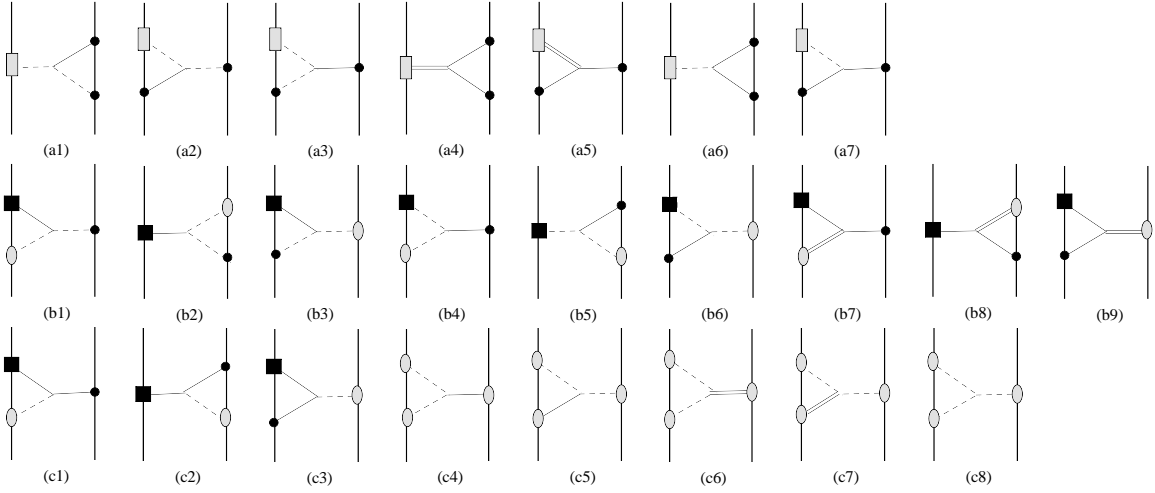}
\caption{The Feynman graphs at one-loop level, i.e.~with cubic 
self-gravitational interaction, which contribute to the NLO cubic-in-spin
interaction at the 4.5PN order for maximally-rotating compact objects.
The graphs should be included together with their mirror images,
i.e.~with the worldline labels $1\leftrightarrow2$ exchanged.
Similar to the nonlinear graphs of two-graviton exchange, these graphs
include all relevant interactions among the spin-induced quadrupole,
octupole, and the mass and spin, and we have here nonlinearities 
originating from gravitons sourced strictly from minimal coupling to the 
worldline, as shown in graphs (c4)-(c8). We also have here cubic vertices 
containing time derivatives, similar to what we have in the NLO odd-parity 
spin-orbit sector \cite{Levi:2010zu,Levi:2015msa,Levi:2015uxa}.}
\label{cubspin1loop}
\end{figure}

As can be seen in figure \ref{cubspin1loop} we have $24$ graphs of cubic
self-interaction in this sector, $6$ of which contain time-dependent
self-interaction, similar to what we have in the odd-parity spin-orbit
sector \cite{Levi:2010zu,Levi:2015msa,Levi:2015uxa}.
Similar to the nonlinear graphs of two-graviton exchange, these graphs
include all relevant interactions among the spin-induced quadrupole,
octupole, and the mass and spin, and we have here nonlinearities
originating from gravitons sourced strictly from minimal coupling to the
worldline, as shown in graphs (c4)-(c8). This sector required using tensor 
one-loop integrals of up to order $5$. 

The graphs in figure \ref{cubspin1loop} are evaluated as follows:
\begin{align}
\text{Fig.~4(a1)}&= -\frac{16}{3}C_{1(BS^3)}\frac{G^2}{r^5}\frac{m_2^2}{m_1^2}
\,\vec{S}_1\cdot \vec{v}_2\times \vec{n}\left[S_1^2-6(\vec{S}_1\cdot\vec{n})^2\right],&\\
\text{Fig.~4(a2)}&= -\frac{3}{2}C_{1(BS^3)}\frac{G^2}{r^5}\frac{m_2}{m_1}
\,\vec{S}_1\cdot \vec{v}_2\times \vec{n}\left[S_1^2-5(\vec{S}_1\cdot\vec{n})^2\right],&\\
\text{Fig.~4(a3)}&= \frac{3}{2}C_{1(BS^3)}\frac{G^2}{r^5}\frac{m_2}{m_1}
\,\vec{S}_1\cdot \vec{v}_1\times \vec{n}\left[S_1^2-5(\vec{S}_1\cdot\vec{n})^2\right],&\\
\text{Fig.~4(a4)}&= -\frac{1}{3}C_{1(BS^3)}\frac{G^2}{r^5}\frac{m_2^2}{m_1^2}\,
\vec{S}_1\cdot\vec{v}_1\times\vec{n}\left[ S_1^2-6 (\vec{S}_1\cdot\vec{n})^2 \right] ,&\\
\text{Fig.~4(a5)}&=-\frac{1}{8}C_{1(BS^3)}\frac{G^2}{r^5}\frac{m_2}{m_1}\,
\vec{S}_1\cdot\vec{v}_1\times\vec{n}\left[S_1^2-5(\vec{S}_1\cdot\vec{n})^2 \big)\right] ,&\\
\text{Fig.~4(a6)}&= \frac{1}{3}C_{1(BS^3)}\frac{G^2}{r^5}\frac{m_2^2}{m_1^2}
\,\vec{S}_1\cdot \vec{v}_2\times \vec{n}\left[S_1^2-6(\vec{S}_1\cdot\vec{n})^2\right],&\\
\text{Fig.~4(a7)}&= \frac{1}{8}C_{1(BS^3)}\frac{G^2}{r^5}\frac{m_2}{m_1}
\,\vec{S}_1\cdot \vec{v}_1\times \vec{n}\left[S_1^2-5(\vec{S}_1\cdot\vec{n})^2\right],&\\
\text{Fig.~4(b1)}&=- \frac{1}{2}C_{1(ES^2)}\frac{G^2}{r^5}\frac{m_2}{m_1}
\,\vec{S}_1\cdot \vec{v}_2\times \vec{n}\left[S_1^2+3(\vec{S}_1\cdot\vec{n})^2\right],&\\
\text{Fig.~4(b2)}&= -8C_{1(ES^2)}\frac{G^2}{r^5}\frac{m_2}{m_1}
\left[\vec{S}_1\cdot \vec{S}_2\times \vec{v}_2\ \vec{S}_1\cdot \vec{n}+
\,\vec{S}_2\cdot \vec{v}_2\times \vec{n}\big(S_1^2-3(\vec{S}_1\cdot\vec{n})^2\big)\right],&\\
\text{Fig.~4(b3)}&= 4C_{1(ES^2)}\frac{G^2}{r^5}
\Big[\,4 \ \vec{S}_1\cdot \vec{S}_2\times \vec{v}_1\ \vec{S}_1\cdot\vec{n}
+\ \vec{S}_1\cdot \vec{S}_2\times \vec{n} \big(\vec{S}_1\cdot \vec{v}_1
-6\vec{S}_1\cdot \vec{n}\ \vec{v}_1\cdot \vec{n}  \big)\nn &\\
&
+\vec{S}_2\cdot \vec{v}_1\times \vec{n}\big(2S_1^2-9(\vec{S}_1\cdot\vec{n})^2\big)
\Big],&\\
\text{Fig.~4(b4)}&= \frac{1}{2}C_{1(ES^2)}\frac{G^2}{r^5}\frac{m_2}{m_1}
\,\vec{S}_1\cdot \vec{v}_1\times \vec{n}\left[S_1^2+3(\vec{S}_1\cdot\vec{n})^2\right]\nn &\\
&-2C_{1(ES^2)}\frac{G^2}{r^4}\frac{m_2}{m_1}
\,\dot{\vec{S}}_1\cdot \vec{S}_1\times \vec{n}\ \vec{S}_1\cdot\vec{n},&\\
\text{Fig.~4(b5)}&= 8C_{1(ES^2)}\frac{G^2}{r^5}\frac{m_2}{m_1}
\Big[\,\vec{S}_1\cdot \vec{S}_2\times \vec{v}_1\ \vec{S}_1\cdot\vec{n}
+\vec{S}_2\cdot \vec{v}_1\times \vec{n}\big(S_1^2-3(\vec{S}_1\cdot\vec{n})^2\big)\Big]\nn &\\
&-4C_{1(ES^2)}\frac{G^2}{r^4}\frac{m_2}{m_1}
\left[\,{\vec{S}}_1\cdot \vec{S}_2\times \vec{n}\ \dot{\vec{S}}_1\cdot\vec{n}
+\dot{\vec{S}}_1\cdot \vec{S}_2\times \vec{n}\ \vec{S}_1\cdot\vec{n}\right],&\\
\text{Fig.~4(b6)}&= 4C_{1(ES^2)}\frac{G^2}{r^5}
\Big[\,2\vec{S}_1\cdot \vec{S}_2\times \vec{v}_1\ \vec{S}_1\cdot\vec{n}
-\vec{S}_1\cdot \vec{S}_2\times \vec{n}
\big(\vec{S}_1\cdot\vec{v}_1-6\vec{S}_1\cdot\vec{n}\ \vec{v}_1\cdot\vec{n}\big)\nn &\\
&
+\vec{S}_2\cdot \vec{v}_1\times \vec{n}\big(2S_1^2-9(\vec{S}_1\cdot\vec{n})^2\big)\Big]
\nn &\\
&
-12C_{1(ES^2)}\frac{G^2}{r^4}
\left[\,{\vec{S}}_1\cdot \vec{S}_2\times \vec{n}\ \dot{\vec{S}}_1\cdot\vec{n}
+\dot{\vec{S}}_1\cdot \vec{S}_2\times \vec{n}\ \vec{S}_1\cdot\vec{n}\right]
,&\\
\text{Fig.~4(b7)}&= -\frac{3}{8}C_{1(ES^2)}\frac{G^2}{r^5}\frac{m_2}{m_1}
\,\vec{S}_1\cdot \vec{v}_1\times \vec{n}\left[S_1^2-5(\vec{S}_1\cdot\vec{n})^2\right],&\\
\text{Fig.~4(b8)}&= 2C_{1(ES^2)}\frac{G^2}{r^5}\frac{m_2}{m_1}
\,\Big[\vec{S}_1\cdot \vec{S}_2\times \vec{v}_2\,\vec{S}_1\cdot\vec{n}
+\vec{S}_2\cdot \vec{v}_2\times \vec{n}\big(S_1^2-3(\vec{S}_1\cdot\vec{n})^2\big)\Big] ,&\\
\text{Fig.~4(b9)}&= \frac{1}{4}C_{1(ES^2)}\frac{G^2}{r^5}
\Big[\,4\vec{S}_1\cdot \vec{S}_2\times \vec{v}_2\ \vec{S}_1\cdot\vec{n}
-2\vec{S}_1\cdot \vec{S}_2\times \vec{n}
\big(\vec{S}_1\cdot\vec{v}_2-3\vec{S}_1\cdot\vec{n}\ \vec{v}_2\cdot\vec{n} \big)\nn &\\
&
+3\vec{S}_2\cdot \vec{v}_2\times \vec{n}\big(S_1^2-5(\vec{S}_1\cdot\vec{n})^2\big)\Big],&\\
\text{Fig.~4(c1)}&= \frac{3}{8}C_{1(ES^2)}\frac{G^2}{r^5}\frac{m_2}{m_1}
\Big[\,\vec{S}_1\cdot \vec{v}_1\times \vec{n}\big(S_1^2-5(\vec{S}_1\cdot\vec{n})^2\big)\Big]\nn &\\
&-C_{1(ES^2)}\frac{G^2}{r^4}\frac{m_2}{m_1}
\left[\,\dot{\vec{S}}_1\cdot \vec{S}_1\times \vec{n}\ \vec{S}_1\cdot\vec{n}\right],&\\
\text{Fig.~4(c2)}&= -2C_{1(ES^2)}\frac{G^2}{r^5}\frac{m_2}{m_1}
\,\Big[\vec{S}_1\cdot \vec{S}_2\times \vec{v}_2 \,\vec{S}_1\cdot\vec{n}
+\vec{S}_2\cdot \vec{v}_2\times \vec{n}\big(S_1^2-3(\vec{S}_1\cdot\vec{n})^2\big)\Big],&\\
\text{Fig.~4(c3)}&= -\frac{1}{4}C_{1(ES^2)}\frac{G^2}{r^5}
\Big[\,4\vec{S}_1\cdot \vec{S}_2\times \vec{v}_1\ \vec{S}_1\cdot \vec{n}
-2\vec{S}_1\cdot \vec{S}_2\times \vec{n}\big(\vec{S}_1\cdot\vec{v}_1
-3\vec{S}_1\cdot\vec{n}\vec{v}_1\cdot\vec{n}\big)\nn&\\
& +3\ \vec{S}_2\cdot \vec{v}_1\times \vec{n}\big(S_1^2-5(\vec{S}_1\cdot\vec{n})^2\big)\Big]\nn&\\
& +C_{1(ES^2)}\frac{G^2}{r^4}
\left[\,{\vec{S}}_1\cdot \vec{S}_2\times \vec{n}\ \dot{\vec{S}}_1\cdot\vec{n}
+\dot{\vec{S}}_1\cdot \vec{S}_2\times \vec{n}\ \vec{S}_1\cdot\vec{n}\right], &\\
\text{Fig.~4(c4)}&= 4\frac{G^2}{r^5}
\left[\,\vec{S}_1\cdot \vec{S}_2\times \vec{v}_2\ \vec{S}_1\cdot\vec{n}
-3\vec{S}_2\cdot \vec{v}_2\times \vec{n}\ (\vec{S}_1\cdot\vec{n})^2\right],&\\
\text{Fig.~4(c5)}&= 4\frac{G^2}{r^5}
\left[\,\vec{S}_1\cdot \vec{S}_2\times \vec{v}_1\ \vec{S}_1\cdot\vec{n}
+\vec{S}_1\cdot \vec{v}_1\times \vec{n}\ \vec{S}_1\cdot\vec{S}_2\right],&\\
\text{Fig.~4(c6)}&= -\frac{1}{2}\frac{G^2}{r^5}
\Big[\,15\vec{S}_1\cdot \vec{S}_2\times \vec{n}
\big(\vec{S}_1\cdot\vec{v}_2-\vec{S}_1\cdot\vec{n}\ \vec{v}_2\cdot\vec{n}\big)\nn &\\
&-\vec{S}_1\cdot \vec{v}_2\times \vec{n}\big(14\vec{S}_1\cdot\vec{S}_2
-12\vec{S}_1\cdot\vec{n}\ \vec{S}_2\cdot\vec{n}\big)+\frac{1}{2}\vec{S}_2\cdot \vec{v}_2\times \vec{n}
\big(29S_1^2-33(\vec{S}_1\cdot\vec{n})^2\big)\Big],&\\
\text{Fig.~4(c7)}&= -\frac{G^2}{r^5}
\Big[4\vec{S}_1\cdot \vec{S}_2\times \vec{v}_1\ \vec{S}_1\cdot\vec{n}
+3\vec{S}_1\cdot \vec{S}_2\times \vec{n}
\big(\vec{S}_1\cdot\vec{v}_1-4\vec{S}_1\cdot\vec{n}\ \vec{v}_1\cdot\vec{n}\big)\nn&\\
&+\vec{S}_2\cdot \vec{v}_1\times \vec{n}\big(S_1^2-6(\vec{S}_1\cdot\vec{n})^2\big)\Big],&\\
\text{Fig.~4(c8)}&= \frac{1}{4}\frac{G^2}{r^5}
\Big[\,4\vec{S_1}\cdot \vec{S}_2\times \vec{v}_1\ \vec{S}_1\cdot\vec{n}
+8\vec{S_1}\cdot \vec{S}_2\times \vec{v}_2\ \vec{S}_1\cdot\vec{n}\nn&\\
&+6\vec{S}_1\cdot \vec{S}_2\times \vec{n}
\big(\vec{S}_1\cdot\vec{v}_1-5\vec{S}_1\cdot\vec{n}\ \vec{v}_1\cdot\vec{n}\big)
+8\vec{S}_1\cdot \vec{S}_2\times \vec{n}
\big(\vec{S}_1\cdot\vec{v}_2-6\vec{S}_1\cdot\vec{n}\  \vec{v}_2\cdot\vec{n} \big)\nn&\\
&
+8\vec{S}_1\cdot \vec{v}_1\times \vec{n}
\big( \vec{S}_1\cdot\vec{S}_2-3\vec{S}_1\cdot \vec{n}\ \vec{S}_2\cdot \vec{n}\big)+
\vec{S}_2\cdot \vec{v}_1\times \vec{n}\big(5S_1^2-9(\vec{S}_1\cdot\vec{n})^2\big)\Big]\nn &\\
&+\frac{G^2}{r^4}
\Big[{\vec{S}}_1\cdot \vec{S}_2\times \vec{n}\ \dot{\vec{S}}_1\cdot\vec{n}
-2{\vec{S}}_1\cdot \dot{\vec{S}}_2\times \vec{n}\ \vec{S}_1\cdot\vec{n}
+\dot{\vec{S}}_1\cdot {\vec{S}}_2\times \vec{n}\ \vec{S}_1\cdot\vec{n}\Big].&
\end{align}

\section{New features from spin dependence of linear momentum}
\label{newfromgauge?}

The formulation of the EFT of a spinning gravitating particle in 
\cite{Levi:2015msa} consisted of an action initially taken in the 
covariant gauge as introduced by Tulczyjew in \cite{Tulczyjew:1959b} 
(later extended to higher-multipoles by Dixon \cite{Dixon:1970zza}). 
Tulczyjew put forward the spin supplementary condition (SSC) given by 
$S_{\mu\nu}p^\nu=0$, which as noted in \cite{Levi:2015msa}, corresponds to 
the choice $e^\mu_0=p^\mu/\sqrt{p^2}$ for the timelike component of the 
worldline tetrad in terms of the linear momentum $p^\mu$. This gauge is 
distinguished among possible covariant gauges, in particular with the 
four-velocity $u^\mu$ (as in $S_{\mu\nu}u^\nu=0$) 
as the only gauge of rotational DOFs for which the existence and uniqueness 
of a corresponding ``center'' for the spinning particle were proven rigorously 
in General Relativity \cite{Schattner:1979vn,Schattner:1979vp}. 

For this reason the formulation in \cite{Levi:2015msa} was made in terms 
of the linear momentum $p^\mu$, rather than the four-velocity $u^\mu$, e.g., 
as in general the former is given by
\be
p_{\mu} = -\frac{\partial L}{\partial u^{\mu}}
= m \frac{u_{\mu}}{\sqrt{u}^2} + {\cal{O}}(RS^2),
\ee
where we recall that the Lagrangian is first constructed with the `spin gauge-invariant' 
variable, as explained in \cite{Levi:2015msa}. 
Therefore the spin-dependent difference between $p_\mu$ and $u_\mu$ would show up, 
as was pointed out in \cite{Levi:2015msa}, as of the NLO of the sector 
cubic in the spins, namely the sector that we are studying in this work.

Let us then find how this new feature transpires in this sector.
Since we are working to cubic order in the spin in this sector,
we should take into account in the linear momentum beyond the leading 
term only the first correction, that is we now consider also
\begin{align} \label{delp}
\Delta p_{\kappa}[S] \equiv p_{\kappa}-\bar{p}_{\kappa}
\simeq \frac{C_{ES^2}}{2m} S^\mu S^\nu 
\left(\frac{2}{u} R_{\mu \alpha \nu \kappa} u^{\alpha}
-\frac{1}{u^3} R_{\mu \alpha \nu \beta} u^{\alpha} u^{\beta} u_{\kappa}
\right),
\end{align}
where we denoted the leading approximation to the linear momentum as 
$\bar{p}_{\kappa}\equiv \tfrac{m}{u} u_{\kappa}$. Let us also note that 
due to eq.~(4.8) of \cite{Levi:2015msa} at this order the expression with 
spin vectors can be used interchangeably as that with the spin tensors.
The appearance of $u^\mu$ in $p^\mu$ itself thus requires further inquiry
only as of the NLO quartic-in-spin sector \cite{Levi:2020lfn}, where it 
was in fact found that this subtlety is still irrelevant until even higher PN orders.

Hence, the part that is linear in the spin in the action of the
spinning particle actually gives rise to a new type of worldline-graviton 
couplings that are cubic in the spin, due to its dependence in the linear 
momentum. We recall that the relevant part of the Lagrangian is given as 
follows \cite{Levi:2015msa}:
\be\label{L_S}
L_{S}=-\frac{1}{2}\hat{S}_{ab}\hat{\Omega}^{ab}_{\text{flat}}
-\frac{1}{2}\hat{S}_{ab} \omega_\mu^{ab} u^{\mu}
-\frac{\hat{S}_{ab}p^b}{p^2}\frac{Dp^a}{D\sigma},
\ee
where the hatted DOFs represent the generic rotational DOFs. 
Therefore the new contributions arise from substituting in 
the linear-in-spin couplings the gauge, 
which we choose here as the canonical gauge, 
formulated in \cite{Levi:2015msa} as
\be
\hat{\Lambda}_{[0]}^{\,a}=\delta_0^a,
\quad \hat{S}^{ab}\left(p_b+p\delta_{0b}\right)=0,
\ee
as well as from the extra term that
enters from minimal coupling, appearing last in eq.~\eqref{L_S}, which
was found in \cite{Levi:2015msa} to be related with the gauge of the 
rotational DOFs, and stands for the Thomas precession as noted in 
section \ref{formulation}. Let us stress again that the subtlety here is 
not about switching from the covariant gauge, but rather about advancing 
from using $u_\nu$ in the basic covariant gauge, to using in it the 
spin-dependent $p_{\nu}$, which is necessary as of this cubic order in 
spins and nonlinear order in gravity. 

Working out explicitly this part of the action in terms of the local spin
variable in the canonical gauge similarly to the derivations 
in \cite{Levi:2015msa}, and keeping only terms that lead to
new cubic-in-spin terms, we obtain here the following contribution:
\be \label{stos3}
L_{S\to S^3}= 
\omega_\mu^{ij} u^\mu \frac{\hat{S}_{ik} p^k p_j}{p\left(p+p^0\right)}
-\omega_\mu^{0i} u^\mu \frac{\hat{S}_{ij}p^j}{p}
+\frac{\hat{S}_{ij}p^i \dot{p}^j}{p\left(p+p^0\right)},
\ee
where in principle all the temporal and spatial indices that are specified 
here are in the locally flat frame.
In order to obtain the new cubic-in-spin couplings we only need to 
substitute in the correction to the linear momentum from eq.~\eqref{delp} 
to linear order, keeping in mind that all of the contributions at the zeroth 
order are taken into account in the Feynman rules from past sectors, 
e.g.~\cite{Levi:2015msa,Levi:2015uxa}, and from section 
\ref{formulation} above.
At this point it becomes clear that the first two terms in 
eq.~\eqref{stos3} give rise to new two-graviton couplings, and that
the last term gives rise to new one-graviton couplings containing 
higher-order time derivatives. 

The resulting new Feynman rules for the 
one-graviton couplings are then:
\begin{align}
\label{eq:stos3A}  
\parbox{12mm}{\includegraphics[scale=0.6]{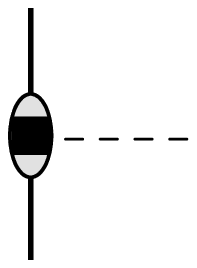}}
 =& \int dt \,\, \Bigg[\frac{C_{\text{ES}^2}}{4m^2}
 S_{i}S_{j}\epsilon_{klm}\Big[\Big(2S_{m} a^{k}  + \dot{S}_{m} v^{k}\Big)
 \Big(A_{l,ij}-A_{j,il}\Big)\Big]\Bigg],\\
\label{eq:stos3phi}   
\parbox{12mm}{\includegraphics[scale=0.6]{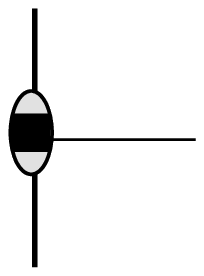}}
 =& \int dt \,\,\Bigg[-\frac{C_{\text{ES}^2}}{2m^2}
 S_{i}S_{j}\epsilon_{klm}\Big[2S_{m}a^k
 \Big(2\big(\phi_{,ij}v^l-\phi_{,il}v^j\big)
 +\delta_{ij}\left(\partial_t\phi_{,l}+\phi_{,ln}v^n\right)\Big)\nn\\
& \qquad \qquad \qquad \qquad 
- \dot{S}_m v^k \Big( 2\phi_{,il}v^j 
- \delta_{ij}\left(\partial_t\phi_{,l} + \phi_{,ln}v^n\right)
+ \delta_{il}\left(\partial_t\phi_{,j} + \phi_{,jn}v^n \right)
 \Big)\Big]\Bigg],
\end{align}
where a black square mounted on a gray oval blob represents this new type of
``composite'' cubic-in-spin worldline couplings. Notice that all these 
rules contain accelerations and even time derivatives of spins, similar to 
the acceleration terms that appear first in the rules for the spin-orbit 
sector \cite{Levi:2015msa}. Note also that at this level the new couplings 
depend linearly on a single Wilson coefficient. 

For the new two-graviton couplings we get the following rules:
\begin{align}
\label{eq:stos3phiA}  
\parbox{12mm}{\includegraphics[scale=0.6]{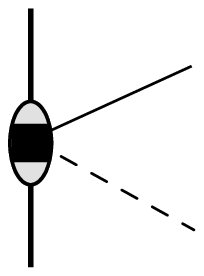}}
 =& \int dt \,\, \Big[\frac{C_{\text{ES}^2}}{2m^2}
S_{i}S_{j}\epsilon_{klm}S_{m}\phi_{,k}\left(
A_{l,ij}-A_{j,il}\right)\Big],\\
\label{eq:stos3phiphi}   
\parbox{12mm}{\includegraphics[scale=0.6]{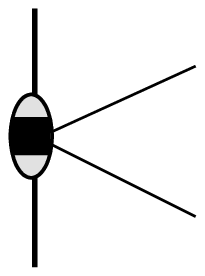}}
 =& \int dt \,\,\left[-\frac{C_{\text{ES}^2}}{m^2}
  S_{i}S_j\epsilon_{klm}S_{m} \phi_{,k}
  \Big(2\big(\phi_{,ij}v^l-\phi_{,il}v^j\big)
   +\delta_{ij}\left(\partial_t\phi_{,l}+\phi_{,ln}v^n\right)\Big)
   \right]. 
\end{align}
Note that the mass ratio together with the Wilson coefficient in these 
new rules for cubic-in-spin couplings indicate that these are truly 
new couplings that cannot be absorbed in the existing ``elementary'' 
octupole operator.

\begin{figure}[t]
\centering
\includegraphics[width=0.8\textwidth]{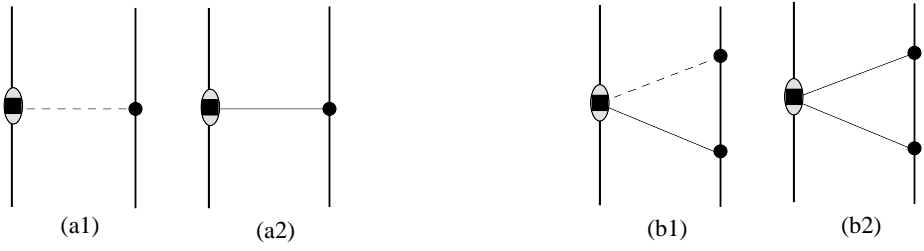}
\caption{The extra Feynman graphs of one- and two-graviton exchange, which
appear at the NLO cubic-in-spin interaction at the 4.5PN order for
maximally-rotating compact objects. 
The graphs should be included together with their mirror images, 
i.e.~with the worldline labels $1\leftrightarrow2$ exchanged. 
These graphs contain a new type of worldline-graviton couplings, which we 
refer to as ``composite'' octupole ones, and obviously yield similar graphs 
to the corresponding ones with the ``elementary'' spin-induced octupole 
couplings in figure \ref{cubspin1g0loop}(a1),(a2) and in figure
\ref{cubspin2g0loop}(b1),(b2).}
\label{cubspinextra0loop}
\end{figure}

These new couplings give rise to $4$ additional graphs as shown in figure
\ref{cubspinextra0loop}, similar to those in figure \ref{cubspin1g0loop} 
(a1), (a2), and in figure \ref{cubspin2g0loop} (b1), (b2). The graphs in 
figure \ref{cubspinextra0loop} are evaluated as follows:
\begin{align}
\text{Fig.~5(a1)}&= -C_{1(ES^2)}\frac{G}{r^3}\frac{m_2}{m_1^2}
\Bigg[2\vec{S}_1\cdot \vec{v}_2\times \vec{a}_1
\left(S_1^2-3\left(\vec{S}_1\cdot \vec{n}\right)^2\right) 
-6\vec{S}_1\cdot \vec{a}_1\times \vec{n}
\vec{S}_1\cdot\vec{v}_2\vec{S}_1\cdot\vec{n}\nn\\
& \qquad \qquad \qquad \quad
-\dot{\vec{S}}_1\cdot \vec{S}_1\times \vec{v}_1 \vec{S}_1\cdot\vec{v}_2 
-\dot{\vec{S}}_1\cdot \vec{v}_1\times \vec{v}_2
\left(S_1^2-3\left(\vec{S}_1\cdot \vec{n}\right)^2\right)\nn\\
& \qquad \qquad \qquad \quad
-3\dot{\vec{S}}_1\cdot \vec{v}_1\times \vec{n}
\vec{S}_1\cdot\vec{v}_2\vec{S}_1\cdot\vec{n}
\Bigg]
,&\\
\text{Fig.~5(a2)}&=
\frac{1}{2} C_{1(ES^2)}\frac{G}{r^3}\frac{m_2}{m_1^2}
\Bigg[6\vec{S}_1\cdot \vec{v}_1\times \vec{a}_1 
\left(S_1^2-2\left(\vec{S}_1\cdot \vec{n}\right)^2\right)
- 2\vec{S}_1\cdot \vec{v}_2\times \vec{a}_1 S_1^2&\nn\\
& \qquad \qquad \qquad \qquad
+6\vec{S}_1\cdot \vec{a}_1\times \vec{n}
\left(S_1^2 \big(\vec{v}_1\cdot\vec{n}-\vec{v}_2\cdot\vec{n}\big)
-2\vec{S}_1\cdot\vec{v}_1\vec{S}_1\cdot\vec{n}\right)&\nn\\
& \qquad \qquad \qquad \qquad
-\dot{\vec{S}}_1\cdot \vec{S}_1\times \vec{v}_1
\left(3\vec{S}_1\cdot\vec{v}_1 - \vec{S}_1\cdot\vec{v}_2 
-3 \vec{S}_1\cdot\vec{n}
\left( \vec{v}_1\cdot\vec{n}-\vec{v}_2\cdot\vec{n}\right)\right)
 &\nn\\
& \qquad \qquad 
+\dot{\vec{S}}_1\cdot \vec{v}_1\times \vec{v}_2 \, S_1^2
+3\dot{\vec{S}}_1\cdot \vec{v}_1\times \vec{n}
\left(S_1^2 \left( \vec{v}_1\cdot\vec{n}-\vec{v}_2\cdot\vec{n}\right)
-2\vec{S}_1\cdot\vec{v}_1\vec{S}_1\cdot\vec{n}\right)
\Bigg]
,&\\
\text{Fig.~5(b1)}&= -2 C_{1(ES^2)}\frac{G^2}{r^5}\frac{m_2^2}{m_1^2}
\,\vec{S}_1\cdot \vec{v}_2\times \vec{n}
\left[S_1^2-3(\vec{S}_1\cdot\vec{n})^2\right],&\\
\text{Fig.~5(b2)}&= C_{1(ES^2)}\frac{G^2}{r^5}\frac{m_2^2}{m_1^2}
\,\Big[3\vec{S}_1\cdot \vec{v}_1\times \vec{n} 
\big(S_1^2-2(\vec{S}_1\cdot \vec{n})^2\big)
-\vec{S}_1\cdot \vec{v}_2\times \vec{n} \,S_1^2\Big].&
\end{align}

\section{The gravitational cubic-in-spin action at the next-to-leading order} 
\label{finalaction}

Let us then put together all the results from sections \ref{Feyncompute} 
and \ref{newfromgauge?} to get the final effective action for this sector. 
This summation includes the values presented above plus similar results 
under the exchange of particle labels $1 \leftrightarrow 2$, where 
$\vec{n}\to-\vec{n}$.
Next, we apply the $4$-vectors identity for $3$ dimensions presented in 
eq.~(3.14) of \cite{Levi:2014gsa}, to further simplify and compress the 
results. As was already noted these results contain higher-order time 
derivatives of both the velocity and the spin, which will be treated 
rigorously at the level of the action, following the procedure shown in 
\cite{Levi:2014sba}, by making variable redefinitions that will remove the
higher-order terms (in complete analogy to the removal of 
redundant/on-shell operators by field redefinitions in effective field 
theories, as was pointed out by one of the authors in \cite{Levi:2014sba}). 

The final result of these steps is then given as follows:
\begin{equation}
L^{\text{NLO}}_{\text{S}^3}=
L^{\text{NLO}}_{\text{S}_1^2\text{S}_2}
+L^{\text{NLO}}_{\text{S}_1^3}+(1\leftrightarrow 2),
\end{equation}
where we have:
\begin{align}
L^{\text{NLO}}_{\text{S}_1^2\text{S}_2}=&\
+\frac{G^2}{r^5}L_{(1)}
+C_{1(ES^2)}\frac{G}{r^4}\frac{1}{m_1}L_{(2)}
+C_{1(ES^2)}\frac{G^2}{r^5}L_{(3)}
+ C_{1(ES^2)}\frac{G^2m_2}{r^5m_1}L_{(4)}
\nn&\\
&
+\frac{G^2}{r^4}L_{(5)}
+C_{1(ES^2)}\frac{G}{r^3}\frac{1}{m_1}L_{(6)}
+C_{1(ES^2)}\frac{G^2}{r^4}L_{(7)}
+C_{1(ES^2)}\frac{G^2m_2}{r^4m_1}L_{(8)}
\nn&\\
&
+C_{1(ES^2)}\frac{G}{r^2}\frac{1}{m_1}L_{(9)}
+C_{1(ES^2)}\frac{G}{r}\frac{1}{m_1}L_{(10)},
\end{align}
with the following pieces:
\begin{align}
L_{(1)} = &
\ \frac{1}{2}\ \vec{S}_1 \cdot\vec{S}_2 \times \vec{n}
\Big(-5\ \vec{S}_1 \cdot\vec{v}_1
+\ \vec{S}_1 \cdot\vec{v}_2
+9\ \vec{S}_1 \cdot\vec{n}\ \vec{v}_1 \cdot\vec{n}
-9\ \vec{S}_1 \cdot \vec{n}\ \vec{v}_2 \cdot \vec{n}
\Big)\nn&\\
&
+9\ \vec{S}_1 \cdot\vec{v}_1 \times \vec{n}\ 
\vec{S}_1 \cdot\vec{n}\ \vec{S}_2 \cdot \vec{n} - \frac{5}{4}\ \vec{S}_2 \cdot\vec{v}_2 \times \vec{n} \big(S_1^2 + 3\big(\vec{S}_1 \cdot \vec{n}\big)^2 \big) 
\nn&\\
&
+\vec{S}_1 \cdot\vec{v}_2 \times \vec{n}
\big( 3\ \vec{S}_1 \cdot\vec{S}_2
-12\ \vec{S}_1 \cdot\vec{n}\ \vec{S}_2 \cdot\vec{n}
\big)
-\frac{1}{4}\ \vec{S}_2 \cdot\vec{v}_1 \times \vec{n}\ \big(7S_1^2 - 27 \big(\vec{S}_1 \cdot\vec{n}\big)^2\big),
\end{align}
\begin{align}
L_{(2)}=&
-3\ \vec{S}_1 \cdot\vec{S}_2 \times \vec{v}_1
\Big(4\ \vec{S}_1 \cdot \vec{v}_1\ 
\vec{v}_2 \cdot \vec{n}
+\vec{S}_1 \cdot\vec{v}_2\ \vec{v}_2\cdot\vec{n} \Big) +\frac{9}{2}\ \vec{S}_2 \cdot \vec{v}_1 \times \vec{v}_2\ S_1^2\ \vec{v}_1 \cdot \vec{n}\nn&\\
&
+3\ \vec{S}_1 \cdot \vec{S}_2 \times \vec{v}_2\Big(4\ \vec{S}_1 \cdot \vec{v}_2\ \vec{v}_1 \cdot \vec{n} + \vec{S}_1 \cdot \vec{n}\big( 2\ \vec{v}_1 \cdot \vec{v}_2 - v_2^2 \big) 
\Big)\nn&\\
&
-3\ \vec{S}_1 \cdot \vec{v}_1 \times \vec{v}_2\ \vec{S}_1 \cdot \vec{S}_2\ \vec{v}_2 \cdot \vec{n} - \frac{15}{2}\ \vec{v}_1 \cdot \vec{v}_2 \times \vec{n}\ \vec{S}_2 \cdot \vec{v}_1 \big( \vec{S}_1 \cdot \vec{n} \big)^2\nn&\\
&
+ \vec{S}_2 \cdot \vec{v}_1 \times \vec{n} \Big(
\frac{15}{2}S^2_1\ \big(v^2_1 - \vec{v}_1 \cdot \vec{v}_2 - 2(\vec{v}_1 \cdot \vec{n})^2
- \vec{v}_1 \cdot \vec{n}\ \vec{v}_2\cdot\vec{n}\big)
\nn&\\
&
 +3\ \vec{S}_1 \cdot \vec{v}_1\big(
-2\ \vec{S}_1 \cdot \vec{v}_1 +2\ \vec{S}_1 \cdot \vec{v}_2
+5\ \vec{S}_1 \cdot \vec{n}\ \vec{v}_1 \cdot \vec{n}\big) - 15\ \vec{S}_1 \cdot \vec{v}_2\ \vec{S}_1 \cdot \vec{n}\ \vec{v}_1 \cdot \vec{n}
 \nn&\\
&
+ \frac{15}{2}\big(\vec{S}_1 \cdot \vec{n}\big)^2\ \big(-v^2_1 + 7\ \vec{v}_1 \cdot \vec{n}\ \vec{v}_2\cdot\vec{n}\big)\Big) +\vec{S}_2 \cdot \vec{v}_2 \times \vec{n} \Big(
\frac{3}{2}\ S^2_1
\big(-5v_1^2 + 4\ \vec{v}_1\cdot\vec{v}_2 - v^2_2
\nn&\\
&
+ 10 (\vec{v}_1 \cdot \vec{n})^2
+ 5\ \vec{v}_1 \cdot \vec{n}\ \vec{v}_2\cdot\vec{n}
\big) + \frac{15}{2}\big(\vec{S}_1 \cdot \vec{n}\big)^2 \big(2v^2_1
- 2\ \vec{v}_1\cdot\vec{v}_2 + v_2^2
- 7\ \vec{v}_1 \cdot \vec{n}\ \vec{v}_2\cdot\vec{n}
\big)\nn&\\
&
+ 3\ \vec{S}_1 \cdot \vec{v}_1\big(
2\ \vec{S}_1 \cdot \vec{v}_1 - \vec{S}_1 \cdot \vec{v}_2
-5\ \vec{S}_1 \cdot \vec{n}\ \vec{v}_1 \cdot \vec{n}
+5\ \vec{S}_1 \cdot\vec{n}\ \vec{v}_2 \cdot\vec{n}
\big)\Big)\nn&\\
&
- 3\ \vec{S}_1 \cdot \vec{v}_1 \times\vec{n}\Big( \vec{S}_1 \cdot \vec{S}_2\ \big(v_1^2 - \vec{v}_1 \cdot \vec{v}_2 -5\ \vec{v}_1 \cdot \vec{n}\ \vec{v}_2 \cdot \vec{n}\big) + \vec{S}_1 \cdot \vec{v}_2\ \vec{S}_2 \cdot \vec{v}_1\nn&\\
&
+ \vec{S}_1 \cdot \vec{v}_1 \big( -\vec{S}_2 \cdot \vec{v}_1 + 5\ \vec{S}_2 \cdot \vec{n}\ \vec{v}_2 \cdot \vec{n} \big)\Big) + 3\ \vec{S}_1 \cdot \vec{v}_2 \times\vec{n}\Big( \vec{S}_1 \cdot \vec{S}_2\ \big(v_1^2 - 5\ \vec{v}_1 \cdot \vec{n}\ \vec{v}_2 \cdot \vec{n} \big)\nn&\\
&
- \vec{S}_1 \cdot \vec{v}_1\ \vec{S}_2 \cdot \vec{v}_1 + 15\ \vec{S}_1 \cdot \vec{v}_2\ \vec{S}_2 \cdot \vec{n}\ \vec{v}_1 \cdot \vec{n}\Big),
\end{align}
\begin{align}
L_{(3)}=&\ \frac{1}{2}\ \vec{S}_1 \cdot\vec{S}_2 \times \vec{v}_1\
\vec{S}_1 \cdot\vec{n} - \frac{1}{2}\ \vec{S}_1 \cdot\vec{S}_2 \times \vec{v}_2\
\vec{S}_1 \cdot\vec{n} + \frac{3}{2}\ \vec{S}_1 \cdot \vec{S}_2 \big( \vec{S}_1 \cdot\vec{v}_1 \times \vec{n} - \vec{S}_1 \cdot\vec{v}_2 \times \vec{n}\big)\nn&\\
&
+\ \vec{S}_1 \cdot\vec{S}_2 \times \vec{n}
\big(-\vec{S}_1 \cdot\vec{v}_1
+\vec{S}_1 \cdot\vec{v}_2
-\frac{3}{2}\ \vec{S}_1 \cdot\vec{n}\ \vec{v}_1 \cdot\vec{n}
+\frac{3}{2}\ \vec{S}_1 \cdot\vec{n}\ \vec{v}_2 \cdot\vec{n}\big)\nn&\\
&
-\frac{3}{4}\ \big(\vec{S}_2 \cdot\vec{v}_1 \times \vec{n} - \vec{S}_2 \cdot\vec{v}_2 \times \vec{n} \big)\big(3S_1^2
-5 (\vec{S}_1 \cdot\vec{n})^2\big),
\end{align}
\begin{align}
L_{(4)}=&
\ 31\ \vec{S}_1\cdot\vec{S}_2\big(\vec{S}_1 \cdot \vec{v}_1 \times\vec{n} - \vec{S}_1 \cdot \vec{v}_2 \times\vec{n}\big)
\nn&\\
&
- 2\Big(\vec{S}_2 \cdot \vec{v}_1 \times \vec{n} - \vec{S}_2 \cdot \vec{v}_2 \times \vec{n} \Big) \big(19 S_1^2 - 21\big(\vec{S}_1 \cdot \vec{n}\big)^2\big)\nn&\\
&
+\vec{S}_1 \cdot \vec{S}_2 \times \vec{n}\big(-41\ \vec{S}_1 \cdot\vec{v}_1 
+41\ \vec{S}_1 \cdot \vec{v}_2 +
63\ \vec{S}_1 \cdot\vec{n}\ \vec{v}_1 \cdot\vec{n} - 
66\ \vec{S}_1 \cdot\vec{n}\ \vec{v}_2 \cdot\vec{n} \big),
\end{align}
\begin{align}
L_{(5)}=&\
3\ \vec{S}_1 \cdot\vec{S}_2 \times \vec{n}\
\dot{\vec{S}}_1 \cdot\vec{n}
-\dot{\vec{S}}_1 \cdot\vec{S}_2
\times \vec{n}\ \vec{S}_1 \cdot\vec{n}
- 2\ \vec{S}_1 \cdot
\dot{\vec{S}}_2 \times \vec{n}\ \vec{S}_1 \cdot\vec{n} + 2\ \vec{S}_1 \cdot \dot{\vec{S}}_1 \times \vec{S}_2,
\end{align}
\begin{align}
L_{(6)}=&\ 
\frac{1}{2}\ \vec{S}_1 \cdot \vec{S}_2 \times \vec{v}_1\ \dot{\vec{S}}_1 \cdot \vec{v}_2 +\frac{1}{2}\ \dot{\vec{S}}_1 \cdot \vec{S}_2 \times \vec{v}_1\ \vec{S}_1 \cdot \vec{v}_2\nn&\\
&
+\frac{1}{2}\ \vec{S}_1 \cdot \vec{S}_2 \times \vec{v}_2\Big(\dot{\vec{S}}_1 \cdot \vec{v}_1 -2\ \dot{\vec{S}}_1 \cdot \vec{v}_2
-3\ \dot{\vec{S}}_1 \cdot \vec{n}\ \vec{v}_1 \cdot \vec{n} -3\ \dot{\vec{S}}_1 \cdot \vec{n}\ \vec{v}_2 \cdot \vec{n}\Big)\nn&\\
&
+\vec{S}_1 \cdot \vec{S}_2 \times \vec{a}_1\ \vec{S}_1 \cdot \vec{v}_2 +\big(\vec{S}_1 \cdot \vec{S}_2 \times \vec{a}_2 + \vec{S}_1 \cdot \dot{\vec{S}}_2 \times \vec{v}_2\big)\big(- \vec{S}_1 \cdot \vec{v}_1 + 3\ \vec{S}_1 \cdot \vec{n}\ \vec{v}_1 \cdot \vec{n} \big)\nn&\\
&
-2\ \vec{S}_1 \cdot \dot{\vec{S}}_2 \times \vec{v}_1\ \vec{S}_1 \cdot \vec{v}_1 +3\ \vec{S}_1 \cdot \vec{v}_2 \times\vec{a}_1\ \vec{S}_1 \cdot \vec{n}\ \vec{S}_2 \cdot \vec{n}\nn&\\
&
+\frac{1}{2}\ \dot{\vec{S}}_1 \cdot \vec{S}_2 \times \vec{v}_2 \Big(\vec{S}_1 \cdot \vec{v}_1 - \vec{S}_1 \cdot \vec{v}_2
-3\ \vec{S}_1 \cdot \vec{n}\ \vec{v}_1 \cdot \vec{n}
-3\ \vec{S}_1 \cdot \vec{n}\ \vec{v}_2 \cdot \vec{n} \Big)\nn&\\
&
- \vec{S}_2 \cdot \vec{v}_1 \times \vec{v}_2\big(\dot{\vec{S}}_1 \cdot \vec{S}_1
-3\ \dot{\vec{S}}_1 \cdot \vec{n}\ \vec{S}_1 \cdot \vec{n}\big)\nn&\\
&
+\big(\vec{S}_2 \cdot \vec{v}_2 \times\vec{a}_2 +\frac{3}{2}\ \vec{S}_2 \cdot \vec{v}_2 \times \vec{a}_1\big)
\big(S_1^2
-3(\vec{S}_1 \cdot \vec{n})^2 \big)\nn&\\
&
+\frac{3}{2}\ \vec{S}_1 \cdot \vec{S}_2 \times \vec{n}
\Big(2\ \vec{S}_1 \cdot\vec{n}\ \vec{a}_1 \cdot \vec{v}_2
+\ \dot{\vec{S}}_1 \cdot\vec{n}\ \vec{v}_1 \cdot \vec{v}_2 -\dot{\vec{S}}_1 \cdot\vec{v}_2\ \vec{v}_1 \cdot \vec{n} \nn&\\
&
+ 5\ \dot{\vec{S}}_1 \cdot \vec{n}\ \vec{v}_1\cdot\vec{n}\ \vec{v}_2 \cdot \vec{n}
\Big) -\frac{1}{2}\big(\vec{S}_2 \cdot \vec{v}_1 \times \vec{a}_2
+\dot{\vec{S}}_2 \cdot \vec{v}_1 \times \vec{v}_2\big)
\big(5 S_1^2 - 9(\vec{S}_1 \cdot \vec{n})^2 \big)\nn&\\
&
+\frac{3}{2}\ \dot{\vec{S}}_1 \cdot \vec{S}_2 \times \vec{n}\Big( -\vec{S}_1 \cdot\vec{v}_2\ \vec{v}_1 \cdot \vec{n} +\vec{S}_1 \cdot\vec{n}\ \vec{v}_1 \cdot \vec{v}_2 + 5\ \vec{S}_1 \cdot\vec{n}\ \vec{v}_1 \cdot \vec{n}\ \vec{v}_2 \cdot \vec{n}\Big)\nn&\\
&
+\frac{3}{2}\ \vec{S}_2 \cdot \vec{v}_1 \times  \vec{n} \Big(
S_1^2\ \vec{a}_1 \cdot \vec{n}
+\dot{\vec{S}}_1 \cdot \vec{S}_1\ (2\vec{v}_1 \cdot \vec{n} + 4\vec{v}_2 \cdot \vec{n}) -10\ \dot{\vec{S}}_1 \cdot \vec{n}\ \vec{S}_1\cdot \vec{n}\ \vec{v}_2 \cdot \vec{n}\Big)\nn&\\
&
+\frac{3}{2}\ \vec{S}_2 \cdot \vec{a}_1 \times \vec{n}
\Big(S_1^2\ \vec{v}_1 \cdot \vec{n}
+S_1^2\ \vec{v}_2 \cdot \vec{n}
-5\ (\vec{S_1}\cdot\vec{n})^2\vec{v}_2\cdot\vec{n} \Big)\nn&\\
&
+\frac{3}{2}\ \vec{S}_1 \cdot \vec{v}_1 \times \vec{n}\big(\vec{S}_1 \cdot \vec{S}_2\ \vec{a}_1 \cdot \vec{n} + 2\vec{S}_1 \cdot \dot{\vec{S}}_2\ \vec{v}_1 \cdot \vec{n}+\dot{\vec{S}}_1 \cdot \vec{S}_2\ \vec{v}_1 \cdot \vec{n} - \dot{\vec{S}}_1\cdot \vec{S}_2\ \vec{v}_2 \cdot \vec{n}\nn&\\
&
- 2\ \dot{\vec{S}}_2\cdot \vec{n}\ \vec{S}_1 \cdot \vec{v}_1 - \vec{S}_1\cdot \vec{n}\ \vec{S}_2 \cdot \vec{a}_1 - \dot{\vec{S}}_1\cdot \vec{n}\ \vec{S}_2 \cdot \vec{v}_1\big)\nn&\\
&
-3\ \vec{S}_1 \cdot \vec{v}_2 \times \vec{n}\ \vec{S}_1 \cdot \vec{S}_2\ \vec{a}_1 \cdot \vec{n}\nn&\\
&
+\vec{S}_2 \cdot \vec{v}_2 \times \vec{n}\Big( \frac{3}{2}(\dot{\vec{S}}_1 \cdot \vec{v}_2\  \vec{S}_1 \cdot \vec{n} + \vec{S}_1 \cdot \vec{v}_2\ \dot{\vec{S}}_1 \cdot \vec{n})\nn&\\
&
-3\ \vec{v}_2 \cdot \vec{n}\ \big(
3\ \dot{\vec{S}}_1 \cdot \vec{S}_1
-5\ \dot{\vec{S}}_1 \cdot \vec{n}\ \vec{S}_1 \cdot \vec{n}\big)\Big)\nn&\\
&
-\frac{3}{2}\ \dot{\vec{S}}_2 \cdot \vec{v}_1 \times \vec{n}
\Big(S_1^2\ \vec{v}_1 \cdot \vec{n} 
-5(\vec{S}_1 \cdot \vec{n})^2 \vec{v}_1 \cdot \vec{n}\Big)\nn&\\
&
+\frac{3}{2}\ \dot{\vec{S}}_1 \cdot \vec{v}_1 \times\vec{n} \Big(\big(\vec{v}_1 \cdot \vec{n} - \vec{v}_2 \cdot \vec{n}\big)\vec{S}_1 \cdot \vec{S}_2 - \vec{S}_1 \cdot \vec{n}\ \vec{S}_2 \cdot \vec{v}_1\Big)\nn&\\
&
+\frac{3}{2}\big(\vec{S}_2 \cdot \vec{a}_2 \times \vec{n}
+\dot{\vec{S}}_2 \cdot \vec{v}_2 \times \vec{n}\big)
\big(-S_1^2\ \vec{v}_1 \cdot \vec{n} 
+2\ \vec{S}_1 \cdot \vec{v}_1\ \vec{S}_1 \cdot \vec{n}
-5(\vec{S}_1 \cdot \vec{n})^2 \vec{v}_1 \cdot \vec{n} \big)\nn&\\
&
+ \frac{3}{2}\ \vec{S}_1 \cdot \vec{a}_1 \times\vec{n} \big( - \vec{S}_1 \cdot \vec{n}\ \vec{S}_2 \cdot \vec{v}_1 + \vec{S}_1 \cdot \vec{S}_2\ \vec{v}_1 \cdot \vec{n} \big),
\end{align}
\begin{align}
L_{(7)}=&\ 
\vec{S}_1 \cdot\vec{S}_2 \times \vec{n}\ \dot{\vec{S}}_1
\cdot\vec{n}
+\dot{\vec{S}}_1
\cdot\vec{S}_2 \times \vec{n}\ \vec{S}_1 \cdot\vec{n},
\end{align}
\begin{align}
L_{(8)}=&\
-4\vec{S}_1 \cdot\vec{S}_2 \times \vec{n}\ \dot{\vec{S}}_1
\cdot\vec{n} -13 \vec{S}_1 \cdot \dot{\vec{S}}_2 \times \vec{n}\ \vec{S}_1 \cdot \vec{n}
-4\dot{\vec{S}}_1
\cdot\vec{S}_2 \times \vec{n}\ \vec{S}_1 \cdot\vec{n},
\end{align}
\begin{align}
L_{(9)}=&\ 
\vec{S}_1 \cdot \dot{\vec{S}}_2 \times \vec{v}_2\ \dot{\vec{S}}_1 \cdot \vec{n} + \dot{\vec{S}}_1 \cdot \dot{\vec{S}}_2 \times \vec{v}_2\ \vec{S}_1 \cdot \vec{n} + \vec{S}_1 \cdot \dot{\vec{S}}_2 \times \vec{a}_1\ \vec{S}_1 \cdot \vec{n}
\nn&\\
&
+\vec{S}_1 \cdot\vec{S}_2 \times \vec{a}_2\ \dot{\vec{S}}_1 \cdot \vec{n} + \dot{\vec{S}}_1 \cdot\vec{S}_2 \times \vec{a}_2\ \vec{S}_1 \cdot \vec{n}-\frac{1}{2}\ \vec{S}_1 \cdot \vec{v}_2\times\vec{n}\ \ddot{\vec{S}}_1 \cdot \vec{S}_2\nn&\\
&
-\frac{1}{2}\ \dot{\vec{S}}_1 \cdot \vec{v}_1 \times\vec{n}\ \vec{S}_1 \cdot \dot{\vec{S}}_2 -\frac{1}{2}\ \ddot{\vec{S}}_1 \cdot \vec{v}_2 \times\vec{n}\ \vec{S}_1 \cdot \vec{S}_2 - \dot{\vec{S}}_1 \cdot \vec{v}_2 \times\vec{n}\ \dot{\vec{S}}_1 \cdot \vec{S}_2\nn&\\
&
-\frac{1}{2}\ \vec{S}_1 \cdot \vec{v}_1 \times\vec{n}\ \dot{\vec{S}}_1 \cdot \dot{\vec{S}}_2 - \vec{S}_2 \cdot \vec{v}_2 \times \vec{n} \big( \ddot{\vec{S}}_1 \cdot \vec{S}_1 + \dot{\vec{S}}_1 \cdot \dot{\vec{S}}_1 \big)\nn&\\
&
-3 \Big(\vec{S}_2 \cdot\vec{a}_2 \times \vec{n}+\dot{\vec{S}}_2 \cdot \vec{v}_2 \times \vec{n}\Big)
\big(\dot{\vec{S}}_1 \cdot \vec{S}_1 - \dot{\vec{S}}_1 \cdot\vec{n}\ \vec{S}_1 \cdot \vec{n}\big)\nn&\\
&
-3\ \dot{\vec{S}}_2 \cdot \vec{v}_1 \times \vec{n}\ \dot{\vec{S}}_1 \cdot \vec{n}\ \vec{S}_1 \cdot \vec{n} -\frac{1}{2}\ \dot{\vec{S}}_2 \cdot \vec{a}_1 \times \vec{n}\Big( 3 \big( \vec{S}_1 \cdot \vec{n} \big)^2 + S^2_1 \Big)\nn&\\
&
- \frac{3}{2}\ \vec{S}_1 \cdot \vec{S}_2 \times \vec{n}\  \ddot{\vec{S}}_1 \cdot \vec{n}\ \vec{v}_2 \cdot \vec{n} + \frac{3}{2}\ \vec{S}_1 \cdot \dot{\vec{S}}_2 \times \vec{n}\ \dot{\vec{S}}_1 \cdot \vec{n}\ \vec{v}_1 \cdot \vec{n}\nn&\\
&
-3\ \dot{\vec{S}}_1 \cdot \vec{S}_2 \times \vec{n}\  \dot{\vec{S}}_1 \cdot \vec{n}\ \vec{v}_2 \cdot \vec{n} -\frac{3}{2}\ \ddot{\vec{S}}_1 \cdot \vec{S}_2 \times \vec{n}\ \vec{S}_1 \cdot \vec{n}\ \vec{v}_2 \cdot \vec{n}\nn&\\
&
+\frac{3}{2}\ \dot{\vec{S}}_1 \cdot \dot{\vec{S}}_2 \times \vec{n}\ \vec{S}_1 \cdot \vec{n}\ \vec{v}_1 \cdot \vec{n},
\end{align}
\begin{align}
L_{(10)}=&\ -\frac{1}{2}\vec{S}_1 \cdot  \dot{\vec{S}}_2 \times \vec{n}\ \ddot{\vec{S}}_1 \cdot \vec{n}
- \dot{\vec{S}}_1 \cdot \dot{\vec{S}}_2 \times \vec{n}\  \dot{\vec{S}}_1 \cdot \vec{n}
-\frac{1}{2} \ddot{\vec{S}}_1 \cdot  \dot{\vec{S}}_2 \times \vec{n}\ \vec{S}_1 \cdot \vec{n},
\end{align}
and also:
\begin{align}
L^{\text{NLO}}_{\text{S}_1^3}=&\
C_{1(ES^2)}\frac{G^2m_2}{r^5m_1}L_{[1]}
+C_{1(ES^2)}\frac{G^2m_2^2}{r^5m_1^2}L_{[2]}
+C_{1(BS^3)}\frac{Gm_2}{r^4m_1^2}L_{[3]}
\nn&\\
&
+C_{1(BS^3)}\frac{G^2m_2}{r^5m_1}L_{[4]}
+C_{1(BS^3)}\frac{G^2m_2^2}{r^5m_1^2}L_{[5]}
+C_{1(ES^2)}\frac{Gm_2}{r^3m_1^2}L_{[6]}
\nn&\\
&
+C_{1(ES^2)}\frac{G^2m_2}{r^4m_1}L_{[7]}
+C_{1(BS^3)}\frac{Gm_2}{r^3m_1^2}L_{[8]}
+C_{1(BS^3)}\frac{Gm_2}{r^2m_1^2}L_{[9]},
\end{align}
with the pieces:
\begin{align}
L_{[1]}=&\ 
\frac{1}{2}(-\vec{S}_1\cdot \vec{v}_1 \times \vec{n}+\vec{S}_1 \cdot \vec{v}_2 \times \vec{n}) \big(S_1^2 - 9\big( \vec{S}_1
\cdot \vec{n}\big)^2 \big),
\end{align}
\begin{align}
L_{[2]}=&\ 
3\Big(\vec{S}_1\cdot \vec{v}_1\times \vec{n}-\vec{S}_1\cdot \vec{v}_2\times \vec{n}\Big)
\left(S_1^2-2(\vec{S}_1\cdot\vec{n})^2\right),
\end{align}
\begin{align}
L_{[3]}=&\ 
\vec{S}_1 \cdot\vec{v}_1 \times \vec{v}_2\ \vec{S}_1 \cdot\vec{v}_2\ \vec{S}_1 \cdot\vec{n}\nn&\\
&
+\vec{S}_1 \cdot\vec{v}_1 \times \vec{n}
\Big(\ \frac{1}{2}\ S_1^2 \big(v_1^2 -2\ \vec{v}_1 \cdot \vec{v}_2 +2v_2^2
- 5\ \vec{v}_1 \cdot \vec{n}\ \vec{v}_2 \cdot \vec{n}
\big)\nn&\\
&
+\vec{S}_1 \cdot \vec{v}_1\big(-\vec{S}_1 \cdot \vec{v}_1
+\vec{S}_1 \cdot\vec{v}_2
+\vec{S}_1 \cdot \vec{n}\ \left(5\ \vec{v}_1 \cdot \vec{n}
-6\ \vec{v}_2 \cdot\vec{n}\right)
\big)
-5\ \vec{S}_1 \cdot\vec{v}_2\
\vec{S}_1 \cdot \vec{n}\ \vec{v}_1 \cdot\vec{n}\nn&\\
&
-\frac{5}{2}(\vec{S}_1 \cdot \vec{n})^2 \big(v_1^2 
-2\ \vec{v}_1 \cdot \vec{v}_2 +2v_2^2
- 7\ \vec{v}_1 \cdot \vec{n}\ \vec{v}_2 \cdot \vec{n}\big)\Big)\nn&\\
&
+\vec{S}_1 \cdot\vec{v}_2 \times \vec{n}
\Big(-\frac{1}{2}S_1^2 \big(v_2^2
- 5\ \vec{v}_1 \cdot \vec{n}\ \vec{v}_2 \cdot \vec{n}
\big) +\frac{5}{2}(\vec{S}_1 \cdot \vec{n})^2 \big(v_2^2
- 7\ \vec{v}_1 \cdot \vec{n}\ \vec{v}_2 \cdot \vec{n}\big)\nn&\\
&
+\vec{S}_1 \cdot \vec{v}_1\big(\vec{S}_1 \cdot \vec{v}_1
-\vec{S}_1 \cdot\vec{v}_2
-\vec{S}_1 \cdot \vec{n}\ \left(4\ \vec{v}_1 \cdot \vec{n}
-5\ \vec{v}_2 \cdot\vec{n}\right)
\big)
+5\ \vec{S}_1 \cdot\vec{v}_2\
\vec{S}_1 \cdot \vec{n}\ \vec{v}_1 \cdot\vec{n}\Big)\nn&\\
&
+ \vec{v}_1 \cdot \vec{v}_2 \times \vec{n} \Big( - \big(  \vec{S}_1 \cdot \vec{n} \big)^2 \Big(\vec{S}_1 \cdot \vec{v}_1 + \frac{5}{2}\ \vec{S}_1 \cdot \vec{v}_2\Big) + \frac{1}{2} S^2_1\ \vec{S}_1 \cdot \vec{v}_2 \Big),
\end{align}
\begin{align}
L_{[4]}=&\ 
\frac{1}{2}\big(\vec{S}_1 \cdot \vec{v}_1
\times \vec{n} - \vec{S}_1 \cdot \vec{v}_2 \times \vec{n} \big) \big(S_1^2 - 5 \big( \vec{S}_1 \cdot \vec{n}\big)^2 \big),
\end{align}
\begin{align}
L_{[5]}=&
-4\ \big(\vec{S}_1\cdot \vec{v}_1 \times \vec{n} - \vec{S}_1 \cdot \vec{v}_2 \times \vec{n}\big) \big(S_1^2
-5\big(\vec{S}_1 \cdot \vec{n}\big)^2 \big),
\end{align}
\begin{align}
L_{[6]}=&\ 
3\Bigg[\Big(\vec{S}_1\cdot \vec{v}_1\times \vec{a}_1-\vec{S}_1\cdot \vec{v}_2\times \vec{a}_1\Big) 
\left(S_1^2-2\left(\vec{S}_1\cdot \vec{n}\right)^2\right)
&\nn\\
& + \vec{S}_1\cdot \vec{a}_1\times \vec{n}
\left(S_1^2 \big(\vec{v}_1\cdot\vec{n}-\vec{v}_2\cdot\vec{n}\big)
-2 \vec{S}_1\cdot\vec{n} 
\big(\vec{S}_1\cdot\vec{v}_1-\vec{S}_1\cdot\vec{v}_2\big)
\right)\Bigg]&\nn\\
&
-\frac{3}{2}\Bigg[\dot{\vec{S}}_1\cdot \vec{S}_1\times \vec{v}_1
\left(\vec{S}_1\cdot\vec{v}_1 - \vec{S}_1\cdot\vec{v}_2 
-\vec{S}_1\cdot\vec{n}
\left( \vec{v}_1\cdot\vec{n}-\vec{v}_2\cdot\vec{n}\right)\right)
 &\nn\\
&
-\dot{\vec{S}}_1\cdot \vec{v}_1\times \vec{v}_2 
\left(S_1^2-2\left(\vec{S}_1\cdot \vec{n}\right)^2\right)
&\nn\\
&
-\dot{\vec{S}}_1\cdot \vec{v}_1\times \vec{n}
\left(S_1^2  \big(\vec{v}_1\cdot\vec{n}-\vec{v}_2\cdot\vec{n}\big)
-2 \vec{S}_1\cdot\vec{n} 
\big(\vec{S}_1\cdot\vec{v}_1-\vec{S}_1\cdot\vec{v}_2\big)
\right)\Bigg],
\end{align}
\begin{align}
L_{[7]}=&\ 
-3\ \dot{\vec{S}}_1 \cdot\vec{S}_1
\times \vec{n}\ \vec{S}_1 \cdot\vec{n},
\end{align}
\begin{align}
L_{[8]}=&
\frac{1}{6}\Big(2\ \vec{S}_1 \cdot \vec{v}_1 \times \vec{v}_2 \big(\dot{\vec{S}}_1 \cdot \vec{S}_1
- 3\ \dot{\vec{S}}_1 \cdot \vec{n}\ \vec{S}_1 \cdot \vec{n} \big)
+ \dot{\vec{S}}_1 \cdot \vec{v}_1 \times \vec{v}_2   \big( S_1^2
-3(\vec{S}_1 \cdot \vec{n} )^2 \big)\nn&\\
&
+\vec{S}_1 \cdot \vec{a}_1 \times \vec{v}_2 \big( S_1^2
-3(\vec{S}_1 \cdot \vec{n} )^2 \big)
+\vec{S}_1 \cdot \vec{v}_1 \times \vec{a}_2
\big( S_1^2-3(\vec{S}_1 \cdot \vec{n} )^2 \big)\nn&\\
&
-6\ \vec{S}_1 \cdot \vec{v}_1 \times \vec{n}\ \Big(\vec{S}_1 \cdot \vec{v}_1\
\dot{\vec{S}}_1 \cdot \vec{n} +\dot{\vec{S}}_1 \cdot \vec{v}_1\ \vec{S}_1 \cdot \vec{n} +\vec{S}_1 \cdot\vec{a}_1\ \vec{S}_1 \cdot \vec{n} \nn&\\
& -\vec{v}_2 \cdot \vec{n} \big(
\dot{\vec{S}}_1 \cdot \vec{S}_1
-5\ \dot{\vec{S}}_1 \cdot\vec{n}\ \vec{S}_1\cdot\vec{n} \big) \Big)\nn&\\
&
-3\ \dot{\vec{S}}_1 \cdot \vec{v}_1 \times \vec{n}\ \Big(2\vec{S}_1 \cdot\vec{v}_1\ \vec{S}_1 \cdot \vec{n} - \vec{v}_2 \cdot \vec{n} \big(S_1^2
-5( \vec{S}_1 \cdot \vec{n} )^2 \big) \Big) \nn&\\
&
-3\ \vec{S}_1 \cdot \vec{a}_1 \times \vec{n}\ \Big(2\vec{S}_1 \cdot\vec{v}_1\ \vec{S}_1 \cdot \vec{n} - \vec{v}_2 \cdot \vec{n} \big(S_1^2
-5( \vec{S}_1 \cdot \vec{n} )^2 \big) \Big)\nn&\\
&
+6\ \vec{S}_1 \cdot \vec{v}_2 \times \vec{n}\Big(\vec{S}_1 \cdot \vec{v}_1\
\dot{\vec{S}}_1 \cdot \vec{n} +
\dot{\vec{S}}_1 \cdot \vec{v}_1\ \vec{S}_1 \cdot \vec{n} + \vec{S}_1 \cdot\vec{a}_1\ \vec{S}_1 \cdot \vec{n} 
\nn&\\
&
- \vec{v}_2 \cdot \vec{n}\big(\dot{\vec{S}}_1 \cdot \vec{S}_1
- 5\ \dot{\vec{S}}_1 \cdot\vec{n}\ \vec{S}_1\cdot\vec{n} \big)\Big)\nn&\\
&
+3\ \dot{\vec{S}}_1 \cdot \vec{v}_2 \times \vec{n}\Big(2\vec{S}_1 \cdot \vec{v}_1\ \vec{S}_1\cdot\vec{n}
-\vec{v}_2 \cdot\vec{n} \big(S_1^2
- 5(\vec{S}_1 \cdot \vec{n})^2 \big)\Big)\nn&\\
&
+3\ \vec{S}_1 \cdot \vec{a}_2 \times \vec{n}\Big( 2\vec{S}_1 \cdot\vec{v}_1\ \vec{S}_1 \cdot\vec{n}
+\vec{v}_1 \cdot \vec{n}\big(S_1^2-5(\vec{S}_1 \cdot \vec{n} )^2\big)\Big)\Big),
\end{align}
\begin{align}
L_{[9]}=&
-\frac{1}{3}\ \vec{S}_1 \cdot \vec{a}_2 \times \vec{n} \big(\dot{\vec{S}}_1 \cdot 
\vec{S}_1
-3\ \dot{\vec{S}}_1 \cdot \vec{n}\ \vec{S}_1 \cdot \vec{n} \big)
-\frac{1}{6}\ \dot{\vec{S}}_1 \cdot \vec{a}_2 \times \vec{n}\big(S_1^2 - 
3\big(\vec{S}_1 \cdot\vec{n}\big)^2\big).
\end{align}

As can be seen in the result above we have grouped together terms 
according to their mass ratios and Wilson coefficients, and the total  
number/order of their higher-order time derivatives. At this stage this 
result is rather bulky, but it is easy to see that after the reduction of 
the higher-order action to an ordinary action by the removal of higher-order 
time derivative terms, we will only be left with such pieces as the first 4 
ones in 
$L_{\text{$S_1^2S_2$}}^{\text{NLO}}$ and the first 5 ones in  
$L_{\text{$S_1^3$}}^{\text{NLO}}$, which becomes significantly more 
compact.
The EOMs can also be derived directly from this higher-order action, 
and then reduced at the level of the EOMs, as was pointed out in 
\cite{Levi:2015msa}.

However, before we will proceed in future work to handle via redefinitions 
the higher-order time derivatives appearing in the cubic-in-spin sector at this 
order, we will need to also take into account all the contributions to the 
action in this sector at this order, which originate from lower-order 
redefinitions of the variables made at lower-order sectors in order to 
remove higher-order time derivatives there, as was shown in detail in 
section 6 of \cite{Levi:2015msa}. First, for example, we recall that we 
have kinematic contributions as noted in eq.~(5.28) of 
\cite{Levi:2015msa}, that are linear in the spin, but have no field 
coupling. Those are required here to NLO as follows:
\be \label{frskin}
L_{\text{kin}}=-\vec{S}\cdot\vec{\Omega}
         - \frac{1}{2} 
         \left(1+\frac{3}{4}v^2\right) \epsilon_{ijk}S_k v^j a^i,
\ee
where $S_{ij}=\epsilon_{ijk}S_k$, and 
$\Omega_{ij}=\epsilon_{ijk}\Omega_k$. 
At LO, e.g., we define the following shift of the positions, $\Delta 
\vec{y}_I$, according to
\begin{equation} \label{positionshift}
\vec{y}_1 \rightarrow \vec{y}_1 + \frac{1}{2 m_1}\vec{S}_1\times\vec{v}_1,
\end{equation}
and similarly for particle 2 with $1\leftrightarrow2$, to remove the
leading accelerations.
Note that as of the NLO linear-in-spin level higher-order time derivatives 
of spin also appear, where it was shown how to generically treat these 
in section 5 of \cite{Levi:2014sba}. Yet, since the leading spin 
redefinition is of higher PN order, terms quadratic in the leading 
redefinition contribute only at the next-to-NNLO (NNNLO) level. Therefore, 
here it is sufficient to consider the redefinition of the spins to linear 
order.

To recap, let us list the additional contributions coming from lower-order 
variable redefinitions that we will have from other sectors.
From position shifts in lower-order sectors we will have:
\begin{enumerate}
\item The LO (1.5PN) position shift in eq.~\eqref{positionshift} 
implemented to linear order on the NLO quadratic-in-spin
(spin1-spin2 + spin-squared) sectors.
\item The above LO position shift implemented to quadratic order on the
Newtonian and LO spin-orbit sectors.
\item The above LO position shift to cubic order implemented on the 
Newtonian sector.
\item The NLO position shift at 2.5PN order in eq.~(6.20) of 
\cite{Levi:2015msa} implemented to linear order on the LO 
quadratic-in-spin sectors.
\item The NLO position shifts at 3PN order in eqs.~(6.30), (6.43) of 
\cite{Levi:2015msa} implemented to linear order on the \textit{shifted} LO 
spin-orbit sector.
\end{enumerate}
The leading redefinition of spin (of 2PN order) in eq.~(6.21) of
\cite{Levi:2015msa} will not contribute to our sector.
From spin redefinitions, i.e.~rotations of the spin, we will have then:
\begin{enumerate}
\item The spin redefinitions at 2.5PN order in eqs.~(6.31), (6.44) of
\cite{Levi:2015msa} implemented to linear order on the LO 
quadratic-in-spin sectors.
\item The spin redefinitions at 3PN order, which were required at the LO 
cubic-in-spin sector \cite{Levi:2014gsa}, 
implemented to linear order on the LO spin-orbit sector. 
\end{enumerate}

In a future publication we will present the full details of these 
redefinitions and the contributions from lower-order sectors, which 
add up to the reduced effective action in this sector.

\section{Conclusions} 
\label{lafin}

In this work we derived for the first time the complete NLO cubic-in-spin 
PN effective action for the interaction of generic compact binaries via 
the self-contained EFT formulation for gravitating spinning objects in 
\cite{Levi:2015msa}, and its extension in this work to the leading sector, 
where gravitational non-linearities are considered at an order in the spins 
that is beyond quadratic. This sector, which enters at the 4.5PN order for 
rapidly-rotating compact objects, completes finite-size effects up to this 
PN order, and is the first sector completed beyond the current state of the 
art for generic compact binary dynamics at the 4PN order. Once again the 
EFT of gravitating spinning objects has enabled a push in the state of 
the art in PN Gravity. Yet the analysis in this work indicates that going 
beyond this sector into the intriguing gray area of table 
\ref{stateoftheart} may become extremely intricate.

We have seen that at this order in spins with nonlinearities in gravity we 
have to take into account additional terms, which arise from a new type of 
worldline couplings, due to the fact that at this order 
the Tulczyjew gauge, which involves the linear momentum, can no longer be 
approximated only in terms of the four-velocity, as the latter 
approximation differs from the linear momentum by a spin-dependent part of 
an order ${\cal{O}}(RS^2)$. The spin-dependent correction gives rise to 
new ``composite'' couplings from the gauge of rotational DOFs. It is 
interesting to consider whether these new couplings have an insightful 
physical interpretation.

As we noted in section \ref{intro} one of the main motivations for us to 
tackle this sector was also to see what happens when we go to a sector at 
order higher than quadratic in the spins and nonlinear in gravity, which 
corresponds to a gravitational Compton scattering with quantum spins of $s 
\ge 3/2$, and to possibly also get an insight on the non-uniqueness of 
fixing its amplitude from factorization when spins of $s\ge5/2$ are 
involved \cite{Arkani-Hamed:2017jhn}. From \cite{Levi:2015msa} and the 
analysis in section \ref{newfromgauge?}, we can see that going to an order 
quintic in the spins, or in the quantum case to $s=5/2$, exactly corresponds 
to where the spin-dependent correction to $p_{\mu}$ in eq.~\eqref{delp} has 
to be taken into account at quadratic order. We will discuss this 
interesting connection between the classical and the quantum levels at a 
future publication.
A general observation that we can clearly make already is that even-parity 
sectors in $l$, see table \ref{stateoftheart}, are easier to handle than 
odd ones. In the quantum context this corresponds to the greater ease of 
dealing with bosons compared to fermions.

Unless all the additional terms from section \ref{newfromgauge?} conspire 
to cancel out eventually, we would obtain an effective action that differs 
from that with the gauge used in lower-spin sectors, involving only the 
four-velocity. 
Even still, it could be that when computing the consequent observable 
quantities, such as the binding energy, or the EOMs, one finds that this 
difference does not matter, and the two gauges are physically equivalent. 
In a forthcoming publication we will present the resulting Hamiltonian, EOMs, 
and gauge-invariant quantities, such as the binding energy, and 
get an answer to these questions, including self-consistency checks of 
the method \cite{Levi:2021xxx}.

At the moment it is not clear whether computations carried out within a 
scattering amplitudes framework can capture all the classical effects 
derived in this paper. The generic results in this work can serve to 
streamline such a framework, as that which was initiated in 
\cite{Chung:2018kqs,Chung:2019duq}, or provide crosschecks for the 
conjectured result for the scattering angle at one-loop level in the 
restricted case of black holes with aligned spins in 
\cite{Guevara:2018wpp}.


\acknowledgments

We thank Yu-tin Huang and Jung-Wook Kim for related discussions.
We are also grateful to Roger Morales and Fei Teng for additional careful 
crosschecks on our results.
The work of ML was supported by the European Research Council under the 
European Union's Horizon 2020 Framework Programme FP8/2014-2020 ``preQFT'' 
grant no.~639729, ``Strategic Predictions for Quantum Field Theories'' project,
and by the European Union's Horizon 2020 research and innovation programme under 
the Marie Sk{\l}odowska-Curie grant agreements No.~847523 `INTERACTIONS' 
and No.~764850 `SAGEX'. We also acknowledge support from the Carlsberg Foundation 
and from the Danish National Research Foundation (DNRF91).



\bibliographystyle{JHEP}
\bibliography{gwbibtex}

\end{document}